\documentclass[twocolumn]{aastex631}

\usepackage[utf8]{inputenc}
\usepackage{graphicx}
\usepackage{indentfirst}
\usepackage{amsmath}
\usepackage{natbib}
\setcitestyle{aysep={}}
\usepackage{rotating}
\usepackage{appendix}

\newcommand{\mbh}{$\mathrm{M}_{\mathrm{BH}}$}
\newcommand{\mbulge}{$\mathrm{M}_{\mathrm{bulge}}$}

\newcommand{\dn}{$\mathrm{D}_\mathrm{n}$4000}

\begin{document}

\title{A Breakdown of the Black Hole - Bulge Mass Relation in Local Active Galaxies}

\author[0000-0003-1055-1888]{Megan R. Sturm}
\affil{Department of Physics, Montana State University, Bozeman, MT 59717, USA}

\author[0000-0001-7158-614X]{Amy E. Reines}
\affil{Department of Physics, Montana State University, Bozeman, MT 59717, USA}

\begin{abstract}    
    
We investigate the relation between black hole (BH) mass and bulge stellar mass for a sample of 117 local ($z \sim 0$) galaxies hosting low-luminosity, broad-line active galactic nuclei (AGN). Our sample comes from \citet{Reines2015}, who found that, for a given {\it total} stellar mass, these AGNs have BH masses more than an order of magnitude smaller than those in early-type galaxies with quiescent BHs. Here we aim to determine whether or not this AGN sample falls on the canonical BH-to-{\it bulge} mass relation by utilizing bulge-disk decompositions and determining bulge stellar masses using color-dependent mass-to-light ratios. We find that our AGN sample remains offset by more than an order of magnitude from the $M_{\rm BH}-M_{\rm bulge}$ relation defined by early-type galaxies with dynamically detected BHs. We caution that using canonical BH-to-bulge mass relations for galaxies other than ellipticals and bulge-dominated systems may lead to highly biased interpretations. This work bears directly to predictions for gravitational wave detections and cosmological simulations that are tied to the local BH-to-bulge mass relations. 
    
\end{abstract}


\section{Introduction}
\label{sec:introduction}

Supermassive black holes (SMBH) are thought to 
live in the center of every massive galaxy \citep{Magorrian1998} and have even been found in dwarf galaxies with masses comparable to the Magellanic Clouds \citep{Reines2013}. SMBHs with masses exceeding $10^9 M_\odot$ have been observed as early as hundreds of millions of years after the Big Bang \citep{Feige2021}, however, it is currently unclear how exactly they formed and grew to such large sizes so quickly. Various seeding mechanisms have been proposed (e.g., stellar remnants from Population III stars \citep{Bond1984} and direct collapse scenarios \citep{Loeb1994, Begelman2006}) and each have their own signatures that should be detectable in the local universe. Semi-analytic models and simulations can predict what the population of SMBHs should look like in the local universe for each seeding mechanism \citep{Volonteri2010, Volonteri2008, Bellovary2019, Ricarte2018}, which we can then compare to observations. 

Clues to BH seeding may come from black hole - host galaxy scaling relations in the low-mass regime \citep{Volonteri2009, Ricarte2018}. Properties including velocity dispersion \citep{Gebhardt2000, Merritt2001}, bulge luminosity \citep{Marconi2003} and bulge mass \citep{Haring2004, McConnell2013, Kormendy2013} have all been found to scale with the mass of the central BH in relatively massive galaxies. Understanding and accurately quantifying these scaling relations across the full range of BH masses, particularly at low masses \citep{Schutte2019,Baldassare2020}, could reveal information about the origin and evolutionary history of SMBHs.

\begin{figure*}
    \centering
    \includegraphics[width=5.5in]{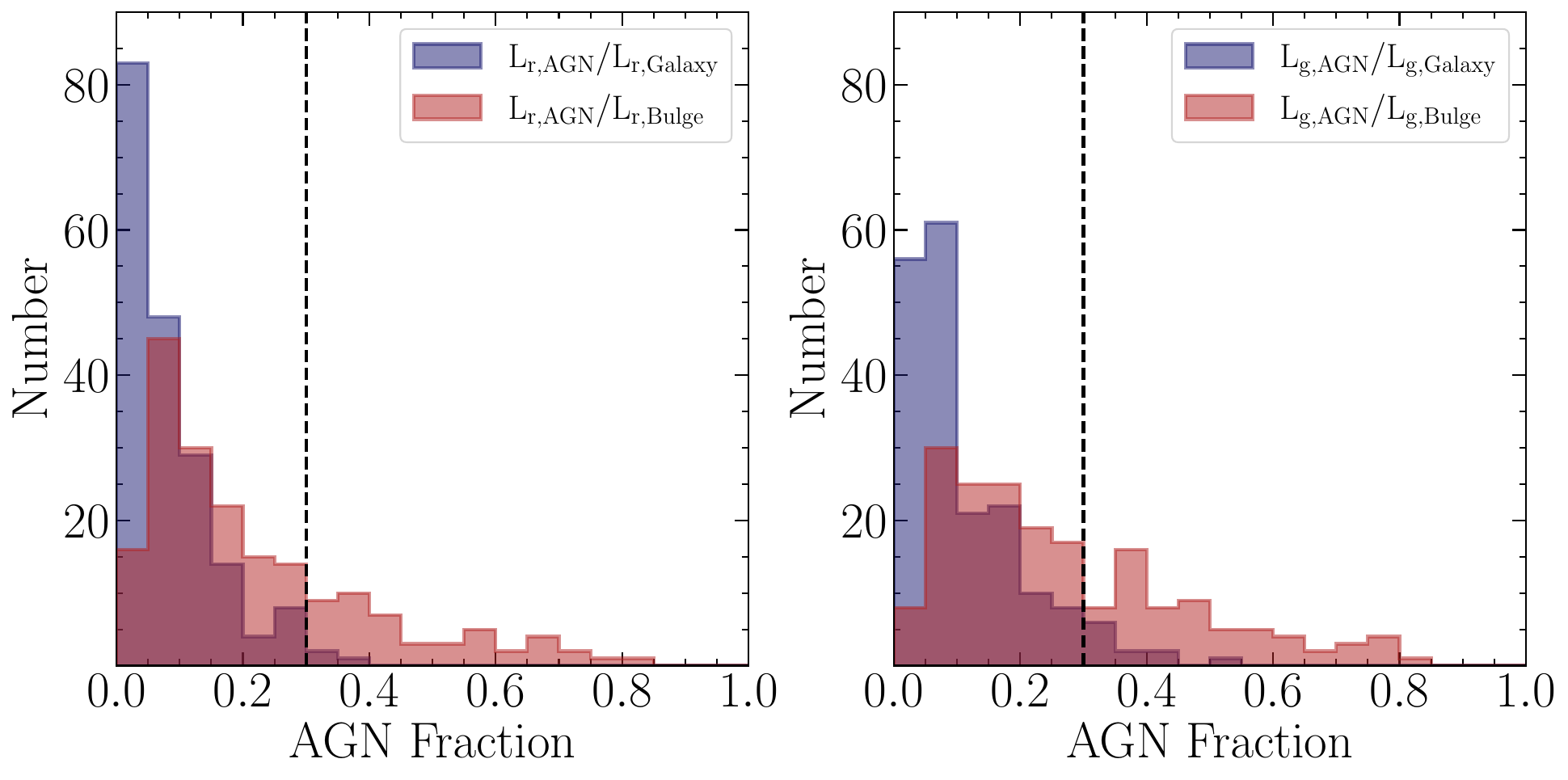}
    \caption{Ratio of AGN to total galaxy luminosity (blue) and bulge luminosity (red) in the $r$- (left panel) and $g$- (right panel) bands. We remove galaxies from our sample where the AGN fraction relative to the bulge in either the $r$- or $g$-band is greater than 30\%, shown by the dashed line in both panels.}
    \label{fig:agn_contribution}
\end{figure*}

Canonical scaling relations have traditionally been derived from dynamical BH mass measurements made from observations biased towards high-mass BHs living in nearby, early-type, inactive galaxies with classical stellar bulges \citep[e.g.,][]{Kormendy2013}. However, these relations are often applied to vastly different types of systems where the relations may not hold, including dwarf or late-type galaxies (where classical bulges may be absent), high redshift galaxies (where stellar bulges, if they exist, cannot be resolved) and active galactic nuclei (AGN, which make up the the vast majority of detected SMBHs). Observations have indeed revealed that some types of galaxies do not follow the same canonical relations defined by early-type galaxies with dynamically detected BHs \citep{Wandel1999, Greene2008, Hu2008, Gadotti2009, Greene2010, Kormendy2011, Sani2011, Graham2014}. 

In particular, \citet{Reines2015} investigated the relationship between BH mass and total stellar mass in a large sample of nearby broad-line AGNs to provide a benchmark for studies that cannot avail themselves of bulge masses and/or dynamical BH masses. \citet{Reines2015} found a clear correlation between BH mass and total stellar mass for the active galaxies with a near-linear slope, similar to that of early-type galaxies with dynamically detected BHs. However, the normalization of the relation defined by the AGNs is more than an order of magnitude lower than that of the comparison sample of early-type galaxies. These findings suggest that early-type galaxies alone provide an incomplete description of the overall population of galaxies hosting central BHs.

Here, we expand upon the work done by \citet{Reines2015}, who noted that the AGN host galaxies on the lower BH-to-{\it total} stellar mass relation could potentially fall on the canonical BH-to-{\it bulge} mass relation if the bulge masses for the AGN hosts were, on average, only 5\% of the total stellar masses. To test this, we use bulge-disk decompositions from \citet{Simard2011} for a subset of the \citet{Reines2015} sample, derive bulge masses, and compare our results to the BH-bulge mass relation in \citet{Kormendy2013}. In Section \ref{sec:sample} we describe the galaxy sample, the bulge-disk decomposition method, and stellar mass calculations. Our results are presented in Section \ref{sec:results} and we conclude with a discussion in Section \ref{sec:conclusion}. As in \citet{Reines2015}, we adopt $\mathrm{H}_0 = 70$ km $\mathrm{s}^{-1}$  $\mathrm{Mpc}^{-1}$ and use SDSS magnitudes.

\section{Galaxy Sample}
\label{sec:sample}

Our primary sample consists of 117 galaxies hosting broad-line AGNs identified by \citet{Reines2015} that also have bulge-disk decompositions in \citet{Simard2011}. As described below, we restrict our broad-line AGN sample to those systems where the AGN luminosity is low compared to the bulge luminosity to help ensure the bulge-disk decompositions are not significantly impacted by the presence of a bright AGN point source.

\begin{deluxetable*}{cccccccccccccccc}
\tablecaption{Sample of 117 Low-Luminosity Broad-line AGN \label{tab:sample}}
\tablewidth{700pt}
\tabletypesize{\scriptsize}
\tablehead{
\colhead{NSA ID} &  \colhead{zdist}  &   \colhead{Model} &  \colhead{$n$} &  \multicolumn{4}{c}{$M_r$} & \colhead{} &
\multicolumn{3}{c}{$g-r$} & \colhead{} &
\multicolumn{3}{c}{log($M/M_\odot$)} \\
\cline{5-8}
\cline{10-12}
\cline{14-16}
\colhead{} &  \colhead{}  &   \colhead{} &  \colhead{} &  \colhead{AGN} &  \colhead{Bulge} &  \colhead{Disk} &  \colhead{Galaxy} & \colhead{} & \colhead{Bulge} &  \colhead{Disk} &  \colhead{Galaxy} & \colhead{} & \colhead{BH} &  \colhead{Bulge} & \colhead{Galaxy}}

\startdata
25955 & 0.04 & n4 & 4.00 & -17.05 & -20.84 & -20.68 & -21.51 & & 0.66 & 0.65 & 0.65 & & 6.21 & 10.45 & 10.72 \\
11183 & 0.03 & free & 4.53 & -16.05 & -17.59 & -19.77 & -19.90 & & 0.69 & 0.61 & 0.62 & & 5.70 & 9.20 & 10.02 \\
7741 & 0.04 & free & 6.41 & -18.57 & -20.12 & -21.17 & -21.51 & & 0.50 & 0.50 & 0.50 & & 6.50 & 9.90 & 10.46 \\
82570 & 0.02 & n4 & 4.00 & -16.89 & -20.69 & -19.84 & -21.10 & & 0.67 & 0.69 & 0.68 & & 6.51 & 10.42 & 10.59 \\
15665 & 0.03 & free & 7.70 & -18.93 & -20.21 & -21.20 & -21.57 & & 0.23 & 0.54 & 0.44 & & 6.74 & 9.49 & 10.45 \\
109525 & 0.05 & n4 & 4.00 & -18.04 & -21.07 & -20.50 & -21.57 & & 0.80 & 0.59 & 0.71 & & 7.25 & 10.79 & 10.89 \\
\enddata

\tablecomments{Sample of 117 
active galaxies with low-luminosity, broad-line AGNs and bulges.
Column 1: NSA identification number. 
Column 2: {\tt zdist} parameter in the NSA, which is based on the SDSS NSA redshift and the peculiar velocity model of \citet{Willick1997}. 
Column 3: Optimal decomposition model from \citet{Simard2011} based on a 1$\sigma$ cutoff on the likelihood. 
Column 4: Bulge Sérsic index if a two-component model is preferred or galaxy Sérsic index if a single-component model is preferred.
Column 5: AGN absolute $r$-band magnitude from \citet{Reines2015}.
Column 6: Bulge absolute $r$-band magnitude from \citet{Simard2011} with the AGN contribution removed. 
Column 7: Disk absolute $r$-band magnitude from \citet{Simard2011}. 
Column 8: Galaxy absolute $r$-band magnitude (from combining AGN-subtracted bulge magnitude and disk magnitude).
Column 8: Bulge ($g-r$) color.   
Column 9: Disk ($g-r$) color.
Column 10: Galaxy ($g-r$) color.
Column 11: Log BH mass from \citet{Reines2015}. 
Column 12: Log bulge stellar mass as calculated in this work. 
Column 13: Log total galaxy stellar mass as calculated in this work. 
Masses are reported as log(M/M$_\odot$). Bulge masses are calculated by subtracting the AGN contribution calculated by \citet{Reines2015} from the bulge light reported in \citet{Simard2011} and using the the mass-to-light ratio from \citet{Zibetti2009}. Disk masses are calculated in the same manner, but without subtracting the AGN contribution. Total galaxy stellar mass is calculated by adding the bulge mass and the disk mass.
For the 14 galaxies that are pure bulge, the disk magnitude and color are reported with a dash. \\ \\
(This table is available in its entirety in machine-readable form.)
}
\end{deluxetable*}

\subsection{Broad-line AGNs}
\label{sec:AGNs}

\citet{Reines2015} started with a large sample of nearby galaxies ($z \le 0.055$) from the NASA-Sloan Atlas (NSA), which provides a reanalysis of the SDSS Data Release 8 (DR8) spectroscopic catalog \citep{Aihara2011} using improved optical photometry \citep{Blanton2011} and spectroscopy \citep{Yan2012,Yan2011}. \citet{Reines2015} performed spectral analysis on $\sim$67,000 galaxies and identified 244 galaxies hosting broad-line AGNs based on the detection of broad H$\alpha$ emission. The broad H$\alpha$ full width at half maximum (FWHM) was restricted to be $\ge 500$~km s$^{-1}$. Additionally, the sources were required to have narrow-line ratios falling in both the AGN region of the [OIII]/H$\beta$ vs.\ [NII]/H$\alpha$ diagram and the Seyfert region of the [OIII]/H$\beta$ vs.\ [SII]/H$\alpha$ diagram. These requirements help ensure a clean sample of broad-line AGNs and eliminate sources of broad H$\alpha$ due to star-formation-related emission \citep[e.g.,][]{Baldassare2016}. Lastly, a handful of sources in which the AGN contributed more than 50\% of the integrated light were removed since the stellar masses for these objects were unreliable. BH masses were estimated using the single-epoch virial mass estimator based on the FWHM and luminosity of broad H$\alpha$ emission and given by equation 5 in \citet{Reines2013}. These BH mass estimates carry uncertainties of $\sim$0.5 dex. 

We cross-match the 244 broad-line AGNs from \citet{Reines2015} with BH mass estimates to the \citet{Simard2011} catalog of 1.2 million galaxies with bulge-disk decompositions, yielding 192 matches.

\begin{figure*}
    \centering
    \includegraphics[width=5.5in]{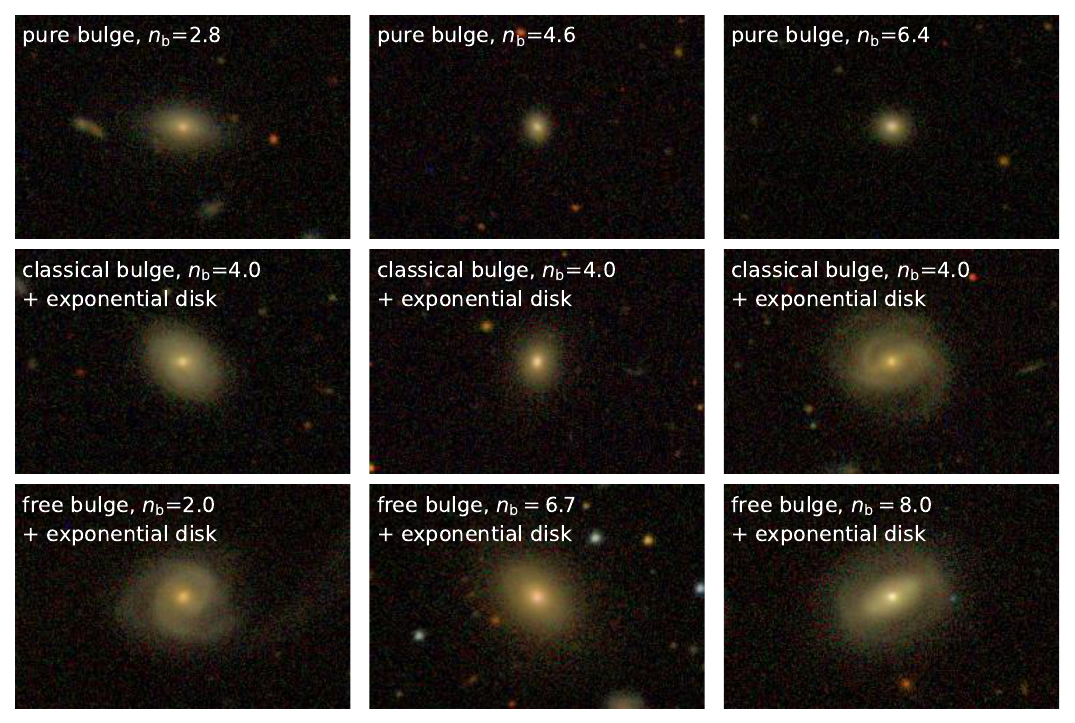}
    \caption{Images of galaxies in our sample from the NSA that were decomposed into a single Sérsic component (top row), a classical bulge + exponential disk (middle row) and a free $n_b$ bulge + exponential disk, determined using a 1$\sigma$ cutoff on the likelihood as reported by \citet{Simard2011}.}
    \label{fig:sample_images}
\end{figure*}

\subsection{Bulge-Disk Decompositions}
\label{sec:decompositions}

\citet{Simard2011} provide catalogs of three different galaxy fitting models for over a million galaxies in the SDSS using Galaxy IMage 2D (GIM2D) to extract bulge and disk components \citep{Simard2002}. The first model fits the entire galaxy with a single Sérsic component with Sérsic index $n_g$. The second model fits the galaxy with two components: an exponential disk (with fixed $n$=1) and a classical bulge (with fixed $n_b=4$). The last model fits the galaxy again with two components, one being an exponential disk (with fixed $n$=1), but in this case, the bulge Sérsic index is a free parameter within the range $0.5<n_b<8$. 

\citet{Simard2011} reports the bulge, disk and total galaxy absolute magnitudes (in the $g$- and $r$-bands) for each fit. They also provide probabilities for each model calculated using $\chi^2$ statistics to indicate the most likely model for each galaxy. Probabilities are provided for each model in relation to more complicated models, i.e. how likely it is that a galaxy requires a two-component model compared to a single-component model or that it needs a free bulge Sérsic index compared to a classical bulge with fixed Sérsic index. We tested a range of cutoff values on the probability for choosing the best model. Increasing the cutoff increases the number of galaxies requiring the most complicated two-component, free $n_b$ model and decreasing it permits the single-component and $n_b=4$ models to begin being utilized. \citet{Simard2011} notes that a lower cutoff extracts a more reliable sample of bulges while a higher cutoff more thoroughly describes overall galaxy properties.

We explore the full range of cutoffs, however, we report results using a 1$\sigma$ cutoff of 0.32 on the probability for choosing the appropriate galaxy model as described in \citet{Simard2011}. This resulted in 119 galaxies best described by the classical bulge + exponential disk model, 58 galaxies best described by a free $n_b$ + exponential disk model and 15 single-component galaxies.  For the single-component galaxies, we set a limit of $n_g>$2 \citep{Kormendy2011}, above which we consider the galaxy to be a pure bulge and below to be a pure disk. One galaxy in our sample was classified as a pure disk (also categorized as an Scd type galaxy in \citealt{Huertas2011}), leaving us with 191 galaxies with bulges. 

The work done by \citet{Simard2011} has generally proven to be robust and is widely used. In their paper, \citet{Simard2011} compares their galaxy Sérsic index and half-light radius values to the New York University Value-Added Galaxy Catalog \citep{Blanton2005}, which also provides Sérsic measurements for SDSS galaxies. They report that, accounting for differences in the two decomposition methods, their decompositions show good agreement with the NYU catalog. Additionally, in recent work on a sample of $\sim$8 million galaxies from the Hyper Suprime-Cam Wide Survey, \citet{Ghosh2023} performs galaxy decompositions using the machine-learning framework Galaxy Morphology Posterior Estimation Network (GaMPEN). They compare their results to the decompositions done by \citet{Simard2011} for the galaxies falling in both samples and find strong correlations between the two measurements for multiple parameters, namely bulge-to-total light ratio, half-light radius and flux.

Furthermore, we find good agreement between the Sérsic index for galaxies requiring a single-component decomposition from \citet{Simard2011} to the Sérsic index found by a single-component fit as reported in the NSA. For the 57,415 galaxies from \citet{Simard2011} that both require a single component decomposition and are also in the NSA, we find a median offset in the two Sérsic indices of 0.12 $\pm$ 1.21 and for the 14 pure bulges in our sample we find a median offset of -0.16 $\pm$ 1.07. Therefore, the Sérsic indices derived through single-component fits in the NSA are generally consistent with those found by \citet{Simard2011}.

As a final check, we use the GIM2D software to reproduce the galaxy models for a subset of our sample using the decomposition parameters chosen based on our 1$\sigma$ cutoff. We find good agreement between the SDSS image and the model for our galaxies and see minimal features in the residuals.

\begin{figure*}
\vspace{-.3cm}
    \centering \includegraphics[width=6in]{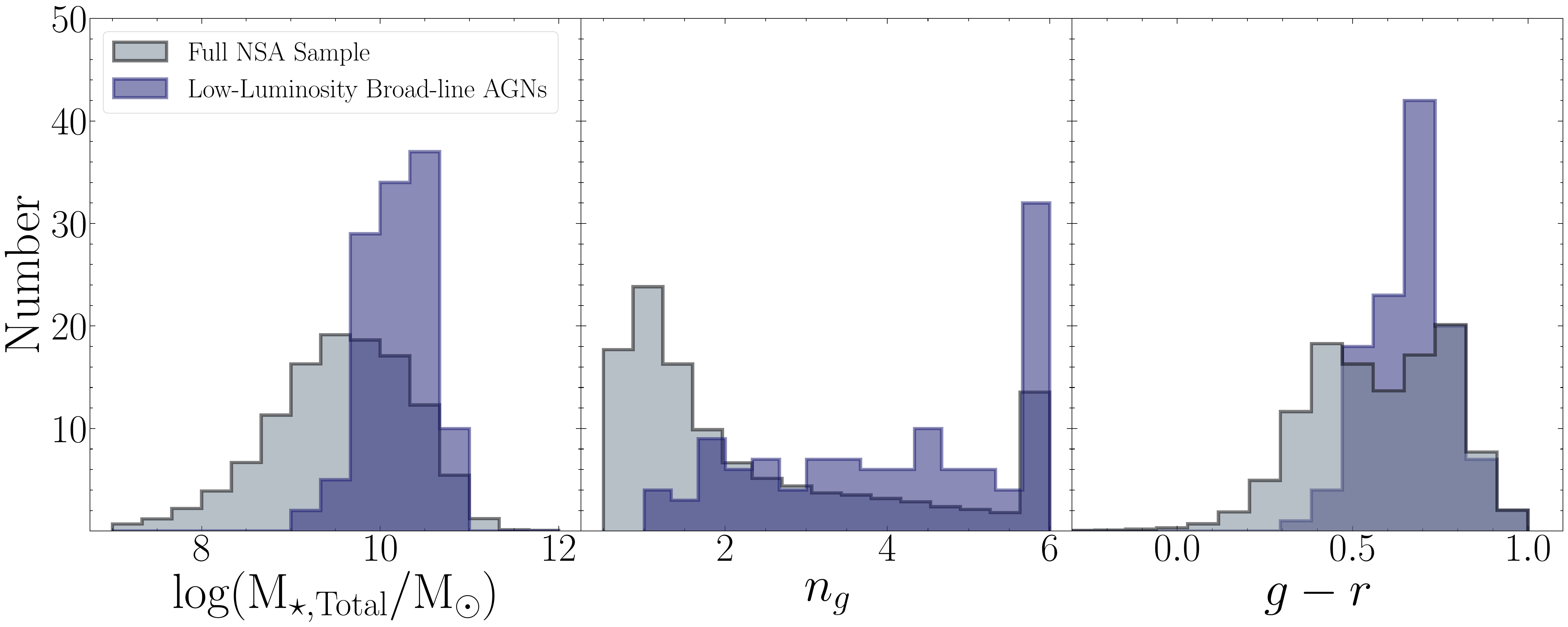}
    \caption{Total stellar mass (left), galaxy Sérsic index (middle) and $g-r$ galaxy color (right) for the 117 active galaxies with low-luminosity, broad-line AGNs in our sample (blue) compared to the full NSA parent sample (gray; normalized to our sample size). All values are as reported in the NSA. Sérsic indices are from single-component fits. We note that the maximum Sérsic index reported by the NSA is 6 causing a false peak. Our sample is biased towards galaxies with higher mass and  higher Sérsic index.
    }
    \label{fig:sample_characterization}
    \vspace{.3cm}
\end{figure*}

\subsection{Restricting to Low-Luminosity AGN}
\label{sec:simard_justification}

Since the decompositions of \citet{Simard2011} do not account for a potential AGN point source at the center of the galaxy, we restrict our sample of broad-line AGNs to those that have a low luminosity relative to the bulge. An unaccounted-for, bright, central point source could skew the Sérsic index towards higher values and the half-light radius towards lower values, (i.e. the galaxy would seem to be more compact than it actually is). This in turn would have the potential to make the decompositions and our subsequent mass measurements unreliable.

Here we dub ``low-luminosity" to indicate that the AGN luminosity is sub-dominant relative to the bulge luminosity in both the $g$-and $r$-bands such that $L_{\rm AGN}/L_{\rm bulge} \le 0.3$. Our threshold of 0.3 is informed by the analysis of \citet{Getachew_Woreta2022} that demonstrates various morphological parameters generally are not significantly impacted until an AGN fraction of at least $\sim$30\%.

Estimated AGN flux densities come from the work of \citet{Reines2015}. They created a mock AGN spectrum consisting of a scaled power law and observed strong emission lines, and then convolved it with the SDSS filter throughput curves. Figure \ref{fig:agn_contribution} shows the distribution of AGN fractions relative to the bulge and total galaxy light. We remove 71 galaxies from our sample where the AGN contribution is above 30\% in either $g$- or $r$-band (or both). {We further address the validity of using the \citet{Simard2011} decompositions for the remaining sample of 120 galaxies in Section \ref{sec:galfit} below.}

Figure \ref{fig:sample_images} shows galaxies in our final sample with different bulge-disk decompositions performed by \citet{Simard2011}. The distribution of total stellar masses, Sérsic indices, and colors for our sample of low-luminosity, broad-line AGNs is shown in comparison with the full NSA parent sample in Figure \ref{fig:sample_characterization}.

\begin{figure*}[!t]
    \centering
    \includegraphics[width=5.5in]{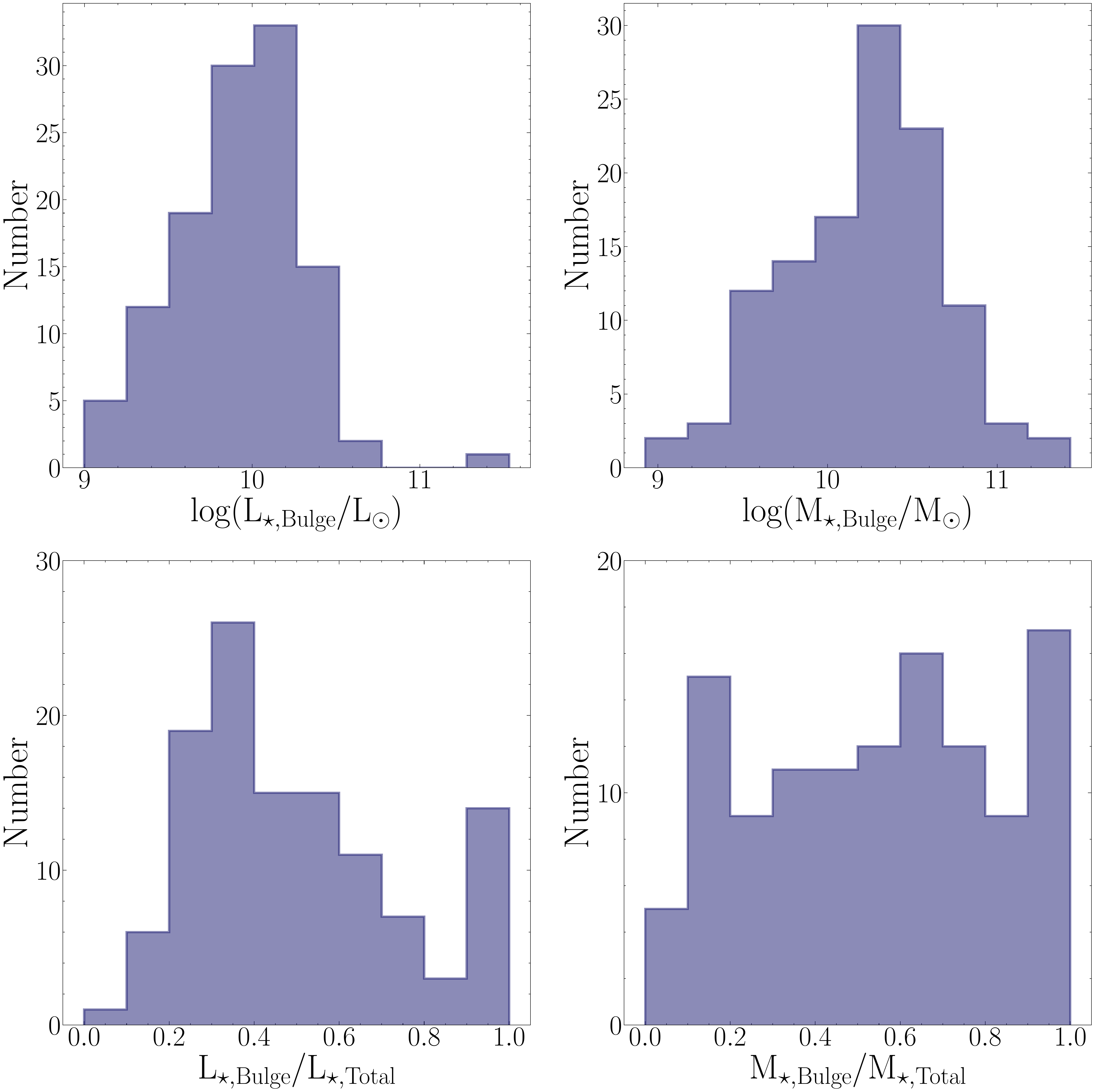}
    \caption{\textbf{Top:} Bulge {$r$-band} luminosity (left) and mass (right). \textbf{Bottom:} Bulge-to-total {$r$-band} luminosity (left) and mass (right) fractions for our sample of low-luminosity, broad-line AGNs. Bulge masses and luminosities are calculated using magnitudes as reported in \citet{Simard2011} with color-depended mass-to-light ratios from \citet{Zibetti2009} and AGN luminosity estimates are from \citet{Reines2015}.}
    \label{fig:bulge_fraction}
\end{figure*}

\subsection{Bulge and Total Stellar Mass Calculations}
\label{sec:mass_calculations}

\begin{figure*}
    \centering
    \includegraphics[width=7in]{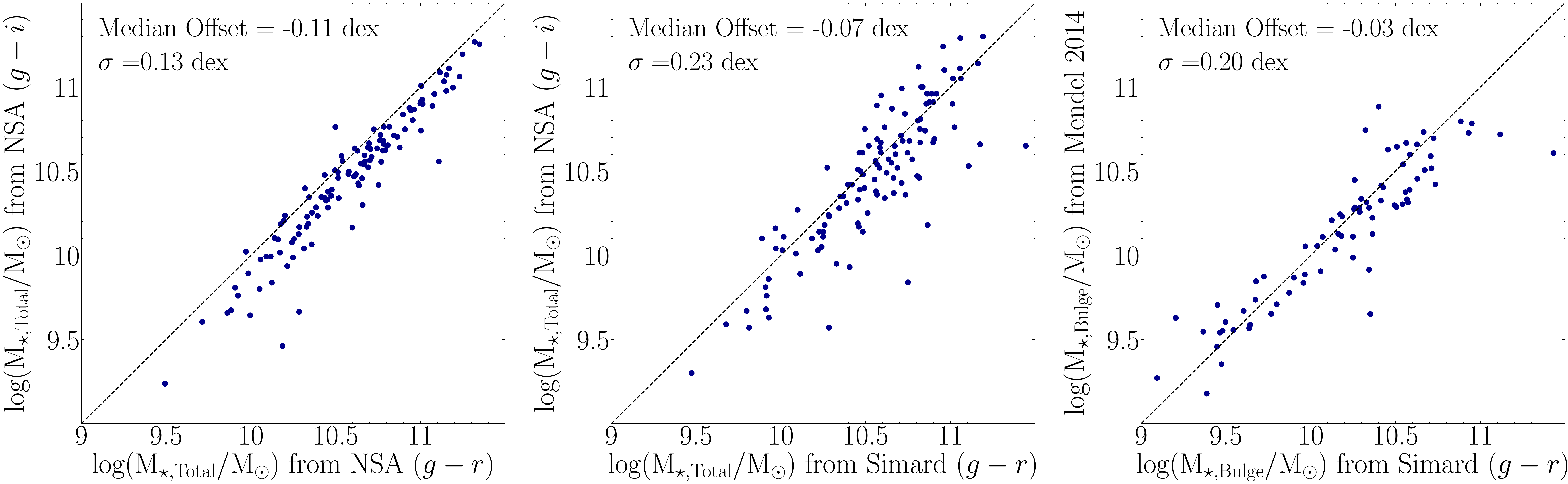}
    \caption{\textbf{Left}: Total stellar masses calculated using different combinations of integrated magnitudes reported in the NSA. Using $g$- and $i$-band data (as in \citealt{Reines2015}) results in a median offset of {-0.11} dex with respect to using $g$-and $r$-band data.
    \textbf{Middle}: Total stellar mass calculated with $g$ and $i$-band data from the NSA again but now versus total stellar masses that we calculate using bulge and disk $g$- and $r$-band data from \citet{Simard2011} and summing the bulge and disk masses. This results in a median offset of -0.07 dex. \textbf{Right:} Bulge mass calculated by \citet{Mendel2014} versus our $g$- and $r$-band-calculated bulge masses. Despite using a different method for determining which decomposition model to use, the median offset is -0.03 dex.}
    \label{fig:mass_comparison}
\end{figure*}

\begin{figure*}
    \centering
    \includegraphics[width=6.5in]{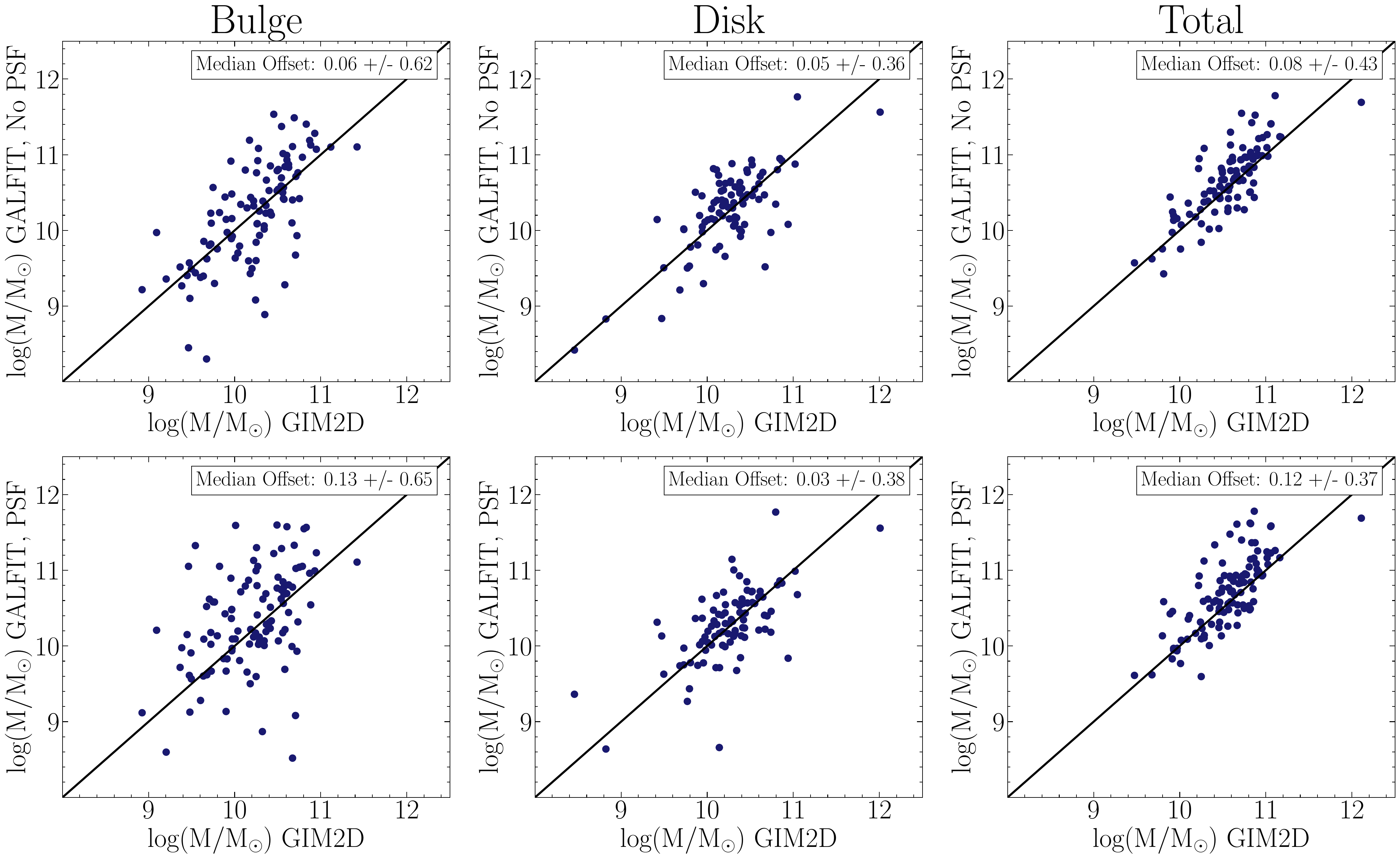}
    \caption{Bulge (left column), disk (middle column) and total (right column) stellar masses calculated using magnitudes determined by two-component (bulge $+$ disk) decompositions from \citet{Simard2011} using the GIM2D software with AGN magnitudes subtracted from the bulge component versus masses derived using magnitudes measured using both two (bulge $+$ disk, top row) and three (bulge $+$ disk $+$ central PSF, bottom row) component fits and the GALFIT software. The black line shows the one-to-one line. We confirm that the two-component fits performed by \citet{Simard2011} are adequate for our sample of low-luminosity AGNs (see \S\ref{sec:galfit}).}
\label{fig:galfit_vs_gim2d_mass}
\end{figure*}

We estimate bulge masses and total stellar masses following a similar procedure as \citet{Reines2015} for consistency, using color-dependent mass-to-light ratios from \cite{Zibetti2009} for the 120 remaining galaxies in our sample. \citet{Simard2011} uses the SDSS spectroscopic redshift, while \citet{Reines2015} adopt the {\tt zdist} parameter provided by the NSA, which is based on the SDSS NSA redshift and the peculiar velocity model of \citet{ Willick1997}. We therefore make a correction to the absolute magnitudes in \citet{Simard2011} to ensure all quantities are calculated consistently using the {\tt zdist} redshift.

Next, we subtract out the AGN contribution to the flux of the bulge and convert disk and bulge magnitudes to disk and bulge masses using the following color-dependent mass-to-light ratio from \citet{Zibetti2009} and a solar $r$-band magnitude of 4.67 from \citet{Bell2003}:

\begin{equation}
    \mathrm{log(M/L}_r) = 1.654(g-r) - 0.840.
    \label{eq:zibetti_g_r}
\end{equation}

\noindent
We adopt total stellar masses as the sum of the bulge mass plus the disk mass. This produces more physically meaningful results than simply converting total galaxy magnitudes provided by \citet{Simard2011} to masses. We found that using the latter method sometimes resulted in a total stellar mass that was less than the calculated bulge mass, which is clearly not physical. Disks and bulges often have very different colors and applying a single color-dependent mass-to-light ratio to the integrated galaxy magnitude would not account for these variations. 

One galaxy had an unreliably high disk mass ($\mathrm{log \: (M_{Disk}/M_{\odot})} \sim 18$) so we remove it from our sample. Additionally, we propagate errors on the magnitudes reported by \citet{Simard2011} through our calculations and find one galaxy with an anomalously high bulge mass error ($\sim$3 dex), so we remove it from our subsequent analysis. The results from our mass calculations are given in Table \ref{tab:sample}. The resulting bulge fractions in both luminosity and mass are shown in Figure \ref{fig:bulge_fraction}. Of this sample of 118 galaxies, 65 are best described by the classical bulge + exponential disk model, 38 are best described by a free $n_b$ + exponential disk model and 15 are single-component galaxies, with one of those being a pure disk galaxy. Therefore, in the end, we have a final sample of 117 galaxies with bulges that we can place on the BH mass - bulge mass relation. 

The stellar masses we calculate are based on $r$-band mass-to-light ratios as a function of $g-r$ color (since \citealt{Simard2011} report magnitudes in the $g$- and $r$-bands), whereas \citet{Reines2015} use $i$-band mass-to-light ratios as a function of $g-i$ color based on integrated photometry in the NSA. To determine the consistency of these methods, we first check how using different bands affects total stellar masses calculated using NSA photometry. Using the corresponding mass-to-light conversions reported in \citet{Zibetti2009}, we calculate total stellar masses using $g$- and $i$-band data (as in \citealt{Reines2015}) as well as $g$- and $r$-band data, both from the NSA, finding that the former produces lower masses with a median offset of 0.11 dex. When we instead compare total stellar masses using $g$- and $i$-band data from the NSA to our total stellar masses calculated using $g$- and $r$-band data from \citet[as bulge mass plus disk mass]{Simard2011}, we find a median offset of 0.07 dex albeit with more scatter. While the offset can be attributed to the different bands, the increased scatter is likely due to using different photometry. Figure \ref{fig:mass_comparison} shows that, overall, our masses are consistent with previously calculated masses in \citet{Reines2015}.

We also compare our masses to those calculated by \citet{Mendel2014}. They provide a catalog of bulge, disk and total stellar mass estimates using SED modelling and the \citet{Simard2011} decompositions. However, we note that they use a slightly different method of distinguishing between galaxies requiring a single or two component fit than we do in this work. 78 of our galaxies have mass estimates in this catalog, and for 10 of these, we used a single component decomposition model while they used a two component model or vice versa. Despite this, we find that our bulge masses are consistent with the ones they calculate, with a median offset of 0.03 $\pm$ 0.20 dex (see Figure \ref{fig:mass_comparison}). Since their calculated bulge masses remain comparable to ours even when a different decomposition model is used, we conclude that our bulge mass estimates are robust.

\begin{figure*}[!t]
    \centering
    \includegraphics[width=6.5in]{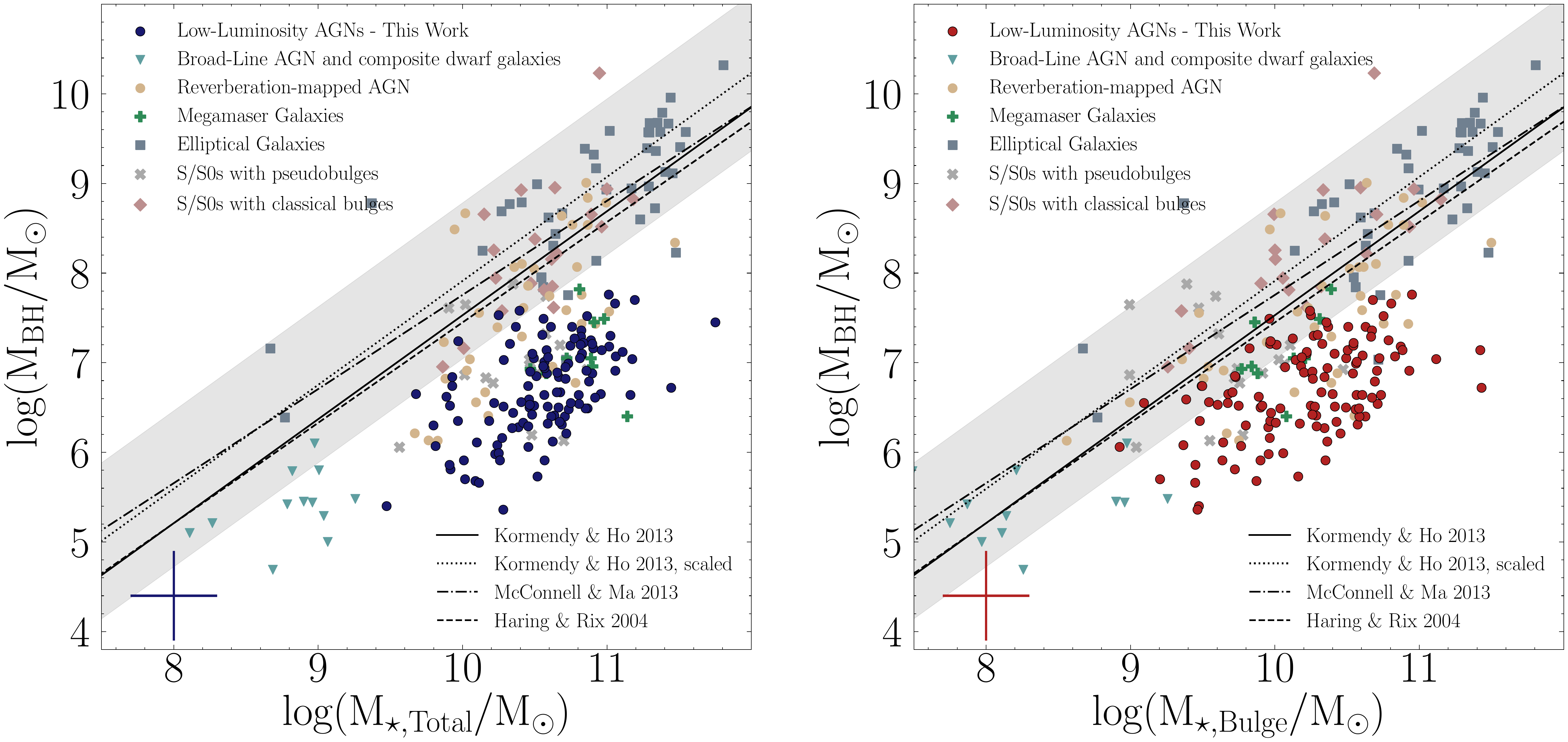}
    \caption{BH mass versus total stellar mass (left) and stellar bulge mass (right).
    The \citet{Kormendy2013}, \cite{Haring2004}, and \citet{McConnell2013} BH-bulge mass relations are shown, with the shaded region indicating the 3$\sigma$ intrinsic scatter of the \citet{Kormendy2013} relation as reported in their paper. Error bars in the bottom left of each plot show the median uncertainties associated with our sample. For comparison, we show the sample of elliptical galaxies and S/S0 galaxies with  classical bulges and pseudobulges, all with dynamical mass measurements. The ellipticals and classical bulges were originally used to derive the \citet{Kormendy2013} relation. We also plot megamaser galaxies from \citet{Lasker2016}, reverberation-mapped AGNs with BH masses from \citet{Bentz2018} and broad-line AGN/composites in dwarf galaxies with bulge masses determined by \citep[][see \S\ref{sec:Additional Objects}]{Schutte2019}.}
    \vspace{.3cm}
    \label{fig:mbh_vs_mgal}
\end{figure*}

\subsection{GALFIT Modeling}\label{sec:galfit}

To further ensure that the presence of low-luminosity AGNs does not impact our results, we re-fit our sample of galaxies using the GALFIT \citep{Peng2010} software both with and without a central point source component in the fitting. The first represents the more thorough method of accounting for the dim central AGN and the latter replicates the fitting performed by \citet{Simard2011}. As described below, we find that the bulge masses do not change significantly by adding a central point source and that we can safely use the decompositions from \citet{Simard2011} despite the fact that they do not include a central point source.

We first create a point-spread function (PSF) for each galaxy image to both convolve with our models and act as the function describing the central AGN point source in the three-component model. We use the parameters provided by SDSS in the corresponding `psField' images and extract a PSF image using the SDSS \texttt{read\_PSF} software utility. We then fit this PSF image in GALFIT using as many Sérsic components as necessary to create an accurate PSF model, which is used in our subsequent modeling.

We follow \citet{Simard2011} in giving apparent and absolute magnitude as:

\begin{equation}
    m = -2.5 \mathrm{log}_{10}(\mathrm{F/t}) - \chi \mathrm{sec}z - m_0 \\
\end{equation}

\begin{equation}
\begin{split}
    M_g = m_g - e_g - DM(z) - k_g \\
    M_r = m_r - e_r - DM(z) - k_r, 
\end{split}
\end{equation}

\noindent
where $\chi$ is the extinction coefficient, sec$z$ is the airmass, $m_0$ is the magnitude zero-point, $e_g$ and $e_r$ are the line-of-sight Galactic extinctions, $DM(z)$ is the distance modulus at redshift $z$ and $k_g$ and $k_r$ are the $k$-corrections. We obtain values for $\chi$, sec$z$ and $m_0$ from the SDSS `tsField' images and values for $e_g$, $e_r$, $k_g$ and $k_r$ from the NSA.

We perform galaxy fits using estimated AGN magnitudes from \citet{Reines2015} and the best-fit parameter values for the bulge and disk from \citet{Simard2011} as the initial parameter guesses for GALFIT. To more closely replicate the simultaneous g- and r-band fits done by \citet{Simard2011}, we first fit the r-band images and use the resulting parameter values for both bulge and disk position angle ($\phi_\mathrm{b}$ and $\phi_\mathrm{d}$), half-light radius ($\mathrm{r}_\mathrm{b}$ and $\mathrm{r}_\mathrm{d}$) and the ellipticity ($\mathrm{e}_\mathrm{b}$ and $\mathrm{e}_\mathrm{d}$) as fixed parameters in the g-band fit. Based on the distribution of $\chi^2$ values from our resulting fits and visual inspection of the images, we consider the GALFIT result successful if the reduced $\chi^2$ value associated with the fit is $<$5. We obtain successful GALFIT models for 95/117 of our galaxies that have bulges. See the Appendix for an example.

We compare the resulting bulge, disk and total stellar masses (calculated using the procedure described above in Section \ref{sec:mass_calculations}) between the two-component GIM2D fit from \citet{Simard2011} and our GALFIT models, both with and without a central point source included as a part of the fit (Figure \ref{fig:galfit_vs_gim2d_mass}). Not including a central point source generally makes the two measurements agree slightly better as expected. However, the median offset in bulge and total mass increases by only $\sim$0.05 dex when we add a PSF component. The median offset between the GALFIT and GIM2d bulge, disk and total stellar masses is small in all cases, $\sim0.04$ dex for the disk and $\sim0.1$ dex for the bulge and total stellar mass. This demonstrates that while there may be a small systematic mass offset between using the two different softwares, the presence of our low-luminosity AGNs does not significantly impact the bulge masses, on average, calculated using a two-component decomposition without a central point source. While the scatter in the bulge masses is $\sim 0.6$ dex, our results on the BH $-$ bulge mass relation would not change significantly as described in more detail in Section \ref{sec:relation}.

\subsection{Additional Objects}
\label{sec:Additional Objects}

For comparison to our low-luminosity, broad-line AGN sample, we also include the objects that are compiled in \cite{Schutte2019} in their study of the BH-bulge mass relation including dwarf galaxies. \citet{Schutte2019} includes 88 galaxies with dynamical BH masses, composed of 79 galaxies from \citet{Kormendy2013} and nine megamaser galaxies from \citet{Lasker2016}. They also include 37 reverberation-mapped AGN with BH masses taken from \citet{Bentz2018}. Lastly, they add 12 broad-line AGNs/composites in dwarf galaxies, two of which are included in our main low-luminosity, broad-line AGN sample. For all of these objects, we follow the methods described in \citet{Schutte2019} to calculate bulge stellar masses in a manner consistent with our low-luminosity, broad-line AGN sample. In other words, all bulge stellar masses are estimated in the most consistent way that is feasible, using color-dependent mass-to-light ratios from \citet{Zibetti2009}. We refer the reader to \citet{Schutte2019} for details on the samples, bulge stellar mass calculations, and BH mass estimates.

\begin{figure}
    \centering
    \includegraphics[width=3in]{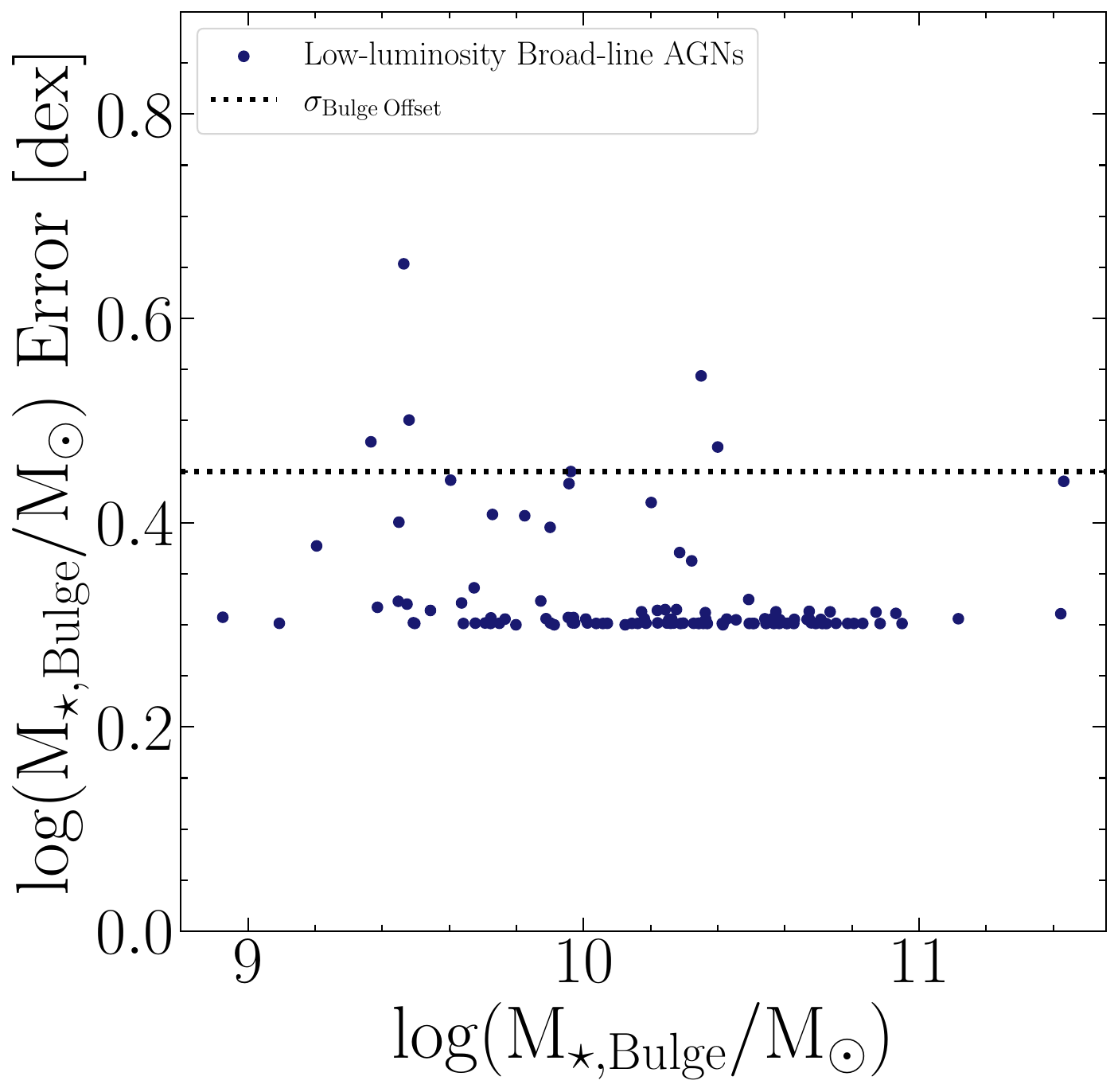}
    \caption{Error in bulge mass versus bulge mass. Errors incorporate uncertainties in the bulge-disk decompositions from \citet{Simard2011} and a 0.3 dex uncertainty from stellar evolution models \citep{Conroy2009}. The dotted line shows the scatter in the distribution of black hole mass versus bulge mass. Only six of our galaxies have individual errors exceeding the scatter in the distribution indicating that the horizontal offset cannot be explained by uncertainties in bulge masses.}
    \label{fig:errors}
\end{figure}

\begin{figure*}
    \centering
    \includegraphics[width=5.5in]{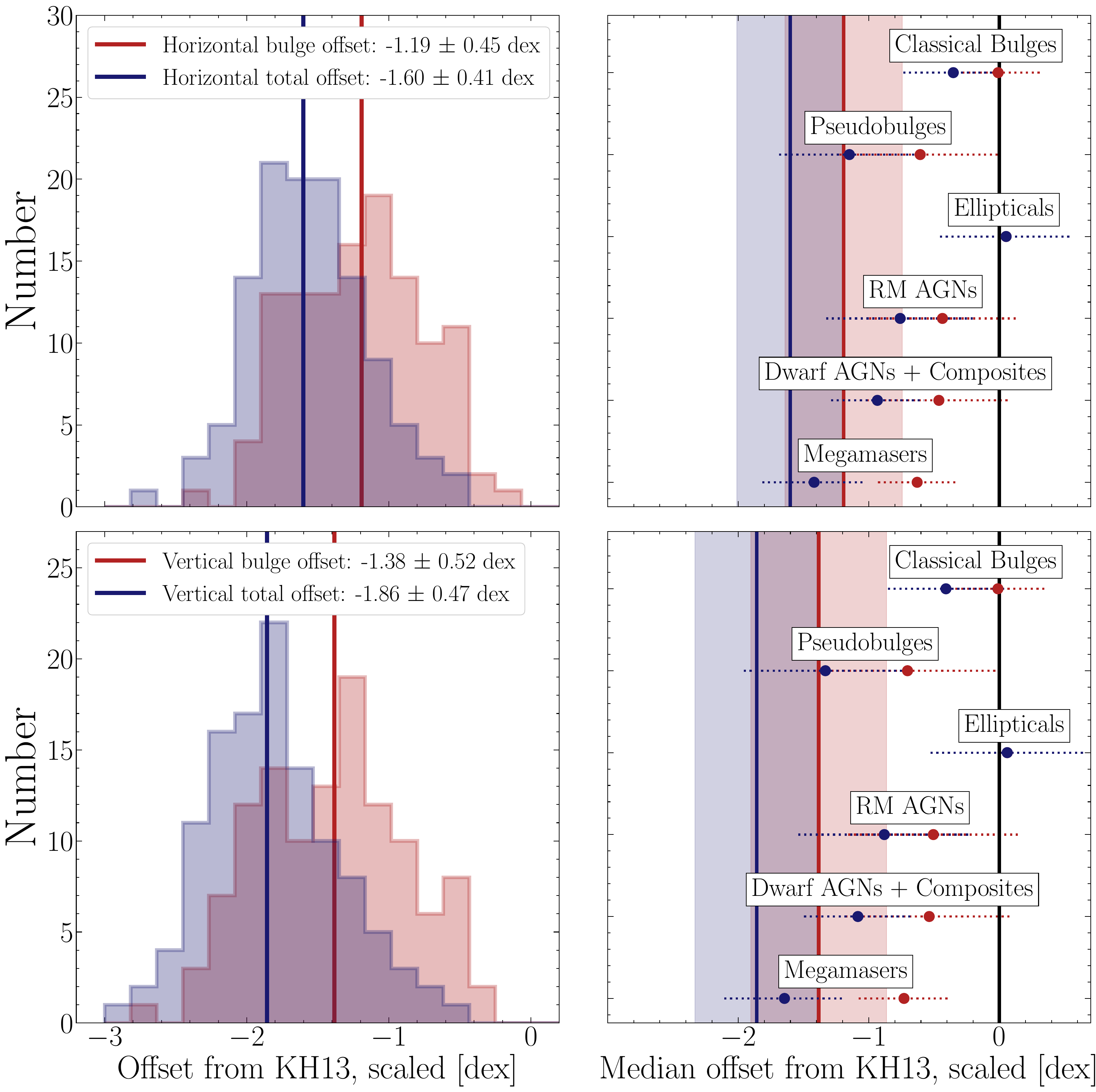}
    \caption{\textbf{Left:} The horizontal (top) and vertical (bottom) offsets of our sample from the \citet{Kormendy2013} \mbh \: versus \mbulge \: relation. We report the 3-$\sigma$-clipped median and standard deviation using the bulge \& total stellar mass. Although using just the bulge mass brings our sample closer to the \citet{Kormendy2013} line, it is still significantly offset. \textbf{Right:} Median horizontal (top) and vertical (bottom) offset from the \citet{Kormendy2013} relation for each of our comparison samples, with blue points indicating the total offset and red points indicating the bulge offset. The vertical lines show the median and the shaded regions show the dispersion for our active galaxy sample. The black vertical line shows zero offset. Dotted error bars represent the 1$\sigma$ scatter in the distribution for each sample. Ellipticals and classical bulges that were originally used to create the relation have a median almost directly on the line, while other types of galaxies tend to fall below, even when the bulge component is extracted.}
    \label{fig:KH_shifts}
\end{figure*}

\begin{figure}
    \centering
    \includegraphics[width=3in]{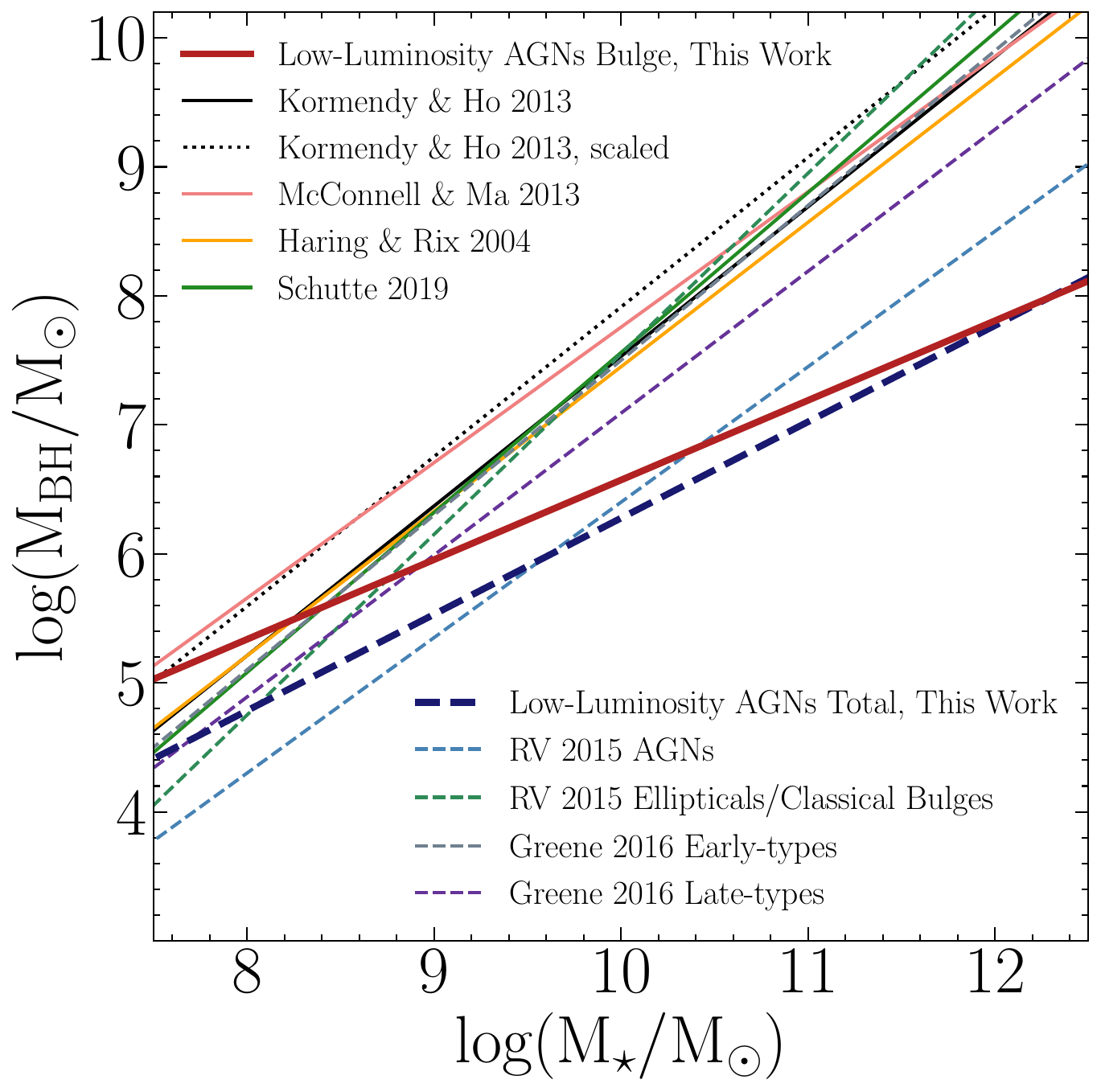}
    \caption{Linear fit of BH mass versus stellar bulge mass (red line) and total stellar mass (blue dashed line) for our sample of low-luminosity, broad-line AGNs in comparison with the canonical \citet{Kormendy2013} BH-bulge mass relation (the scaled relation is shown as a dotted line), the \citet{Reines2015} relation (labeled as RV 2015 AGNs) for BH mass versus total stellar mass for AGNs and multiple other relations from the literature. We indicate BH-bulge mass relations with solid lines and BH-total mass relations with dashed lines (see \S\ref{sec:fit}.)}
    \label{fig:linregress}
\end{figure}

\section{Results}
\label{sec:results}

\subsection{BH-Bulge Mass Relation}\label{sec:relation}

We plot BH mass as a function of both total stellar mass and bulge mass in Figure \ref{fig:mbh_vs_mgal}. The BH mass -  total stellar mass relation was originally explored in \citet{Reines2015}, where they found that the AGN host galaxies fell on average an order of magnitude below early-type galaxies with dynamically detected BHs. Here we intend to determine if using bulge mass rather than total stellar mass brings the AGN host galaxies in line with the early-type galaxies that define the canonical BH mass - bulge mass relations.

For consistency with \citet{Reines2015}, we primarily compare our results to the \citet{Kormendy2013} BH-bulge mass relation (equation 10 in their paper). As noted in \cite{Reines2015}, there is a difference in the assumed IMFs between \citet{Kormendy2013} and \citet{Zibetti2009} that results in systematically higher masses in the former by 0.33 dex, so we scale the \citet{Kormendy2013} relation accordingly since we have adopted masses using the \citet{Zibetti2009} relations throughout this work.

For each of the galaxies in our low-luminosity, broad-line AGN sample we calculate the horizontal shift from the canonical relation (i.e., the offset in bulge mass at a given BH mass). The median horizontal offset from the \citet{Kormendy2013} line is -1.60 $\pm$ 0.41 dex for total stellar mass and -1.19 $\pm$ 0.45 dex for the bulge mass. Extracting the bulge component of the stellar mass shifts the median of the sample $\sim$ 0.41 dex closer to the \cite{Kormendy2013} relation and has a similar scatter, however the sample remains significantly offset from the canonical relation. Similarly, the active galaxies fall below the \citet{Kormendy2013} relation with a median vertical offset of -1.86 $\pm$ 0.48 dex for the total stellar mass and -1.38 $\pm$ 0.52 for the bulge mass. Moreover, at a given bulge mass, there is a large spread in BH masses.

The errors reported above are the 1$\sigma$ scatter in the distribution of offsets. These values are larger than the systematic errors associated with individual points, indicating the offsets are significant. For the vertical offset, the scatter in the distribution is larger than the systematic uncertainties in single epoch virial BH masses ($\sim$0.5 dex). For the horizontal offset, we find individual bulge mass errors by propagating systematic uncertainties through our calculations (see \S\ref{sec:mass_calculations}), resulting in a median error of $\sim$0.06 dex. In most cases, our stellar and bulge mass errors are therefore dominated by uncertainties in stellar evolution, expected to be $\sim$0.3 dex \citep{Conroy2009}. Adding the individual errors and the uncertainties in stellar evolution in quadrature, the scatter in the distribution of horizontal offsets for our sample is greater than the uncertainties for all but six galaxies, shown in Figure \ref{fig:errors}. This indicates that errors in bulge mass estimates cannot account for the discrepancy between our sample of low-luminosity, broad-line AGNs and the canonical BH-bulge mass relation.

We note that changing the cutoff probability for model selection (see \S\ref{sec:decompositions}) in \citet{Simard2011} does not significantly change the bulge offsets from the canonical relation. When the cutoff was set to zero and all galaxies were described with pure Sérsic profiles, the median bulge mass (horizontal) offset was -1.27 $\pm$ 0.47 dex. Increasing the cutoff to one, with all galaxies being described by the two-component, free $n_b$ model, returns a median bulge offset of -1.16 $\pm$ 0.48 dex. In both cases, the AGN sample is still significantly offset from the canonical relation.

\begin{figure*}[!t]
    \centering
    \includegraphics[width=6.4in]{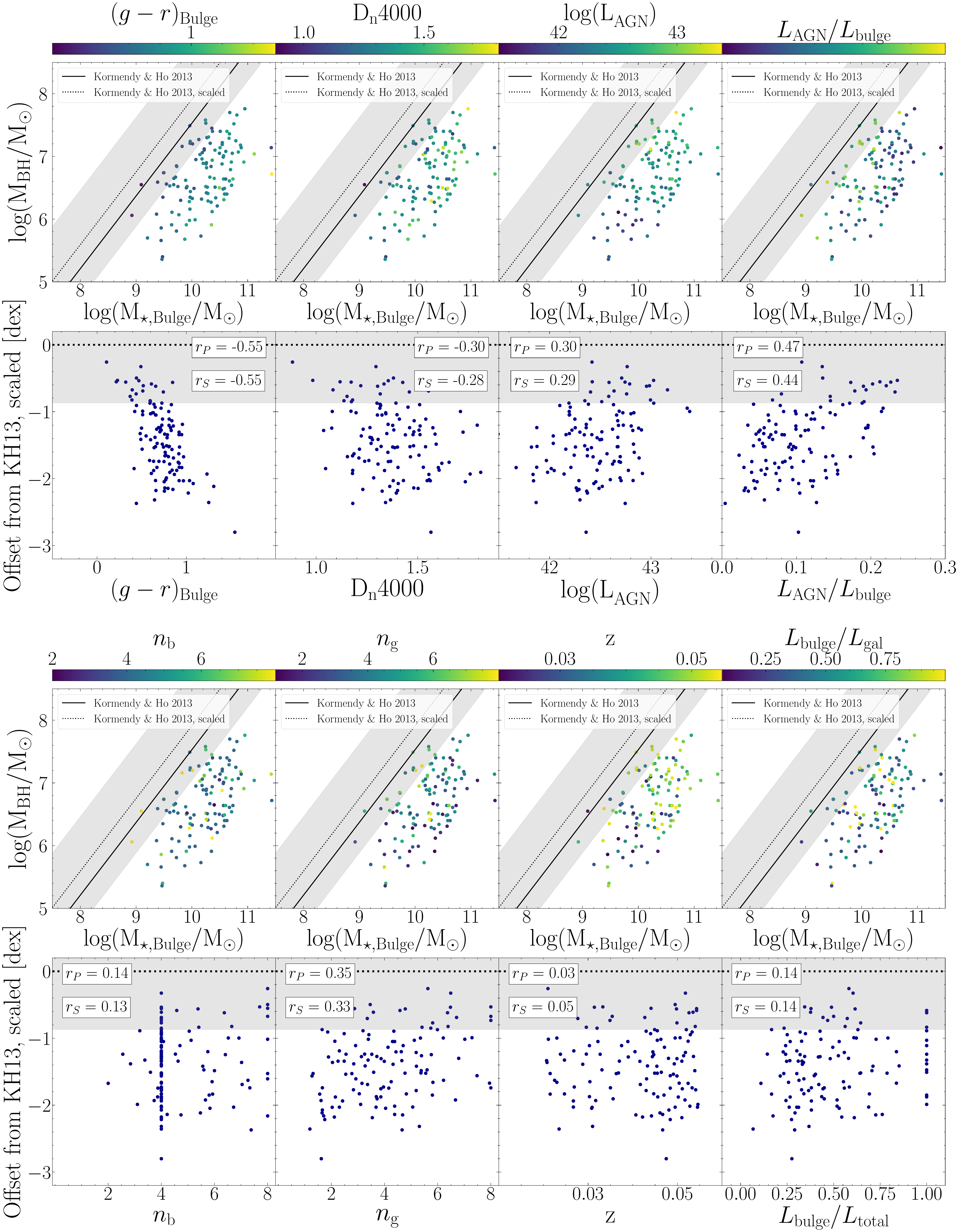}
    \caption{Bulge color, \dn, AGN $r$-band luminosity, AGN-to-bulge $r$-band luminosity ratio, bulge and galaxy Sérsic indices, redshift and bulge-to-total $r$-band luminosity ratio versus the offset in BH mass from the \citet{Kormendy2013} relation. Negative offset corresponds to BHs with observed masses less than predicted. The dotted line shows zero offset. In all plots, the shaded region indicates the 3$\sigma$ intrinsic scatter of the \citet{Kormendy2013} scaled relation as reported in their paper. We find no strong correlations with galaxy properties and their offset from the \citet{Kormendy2013} relation.}
    \label{fig:galaxy_properties}
\end{figure*}

Furthermore, when using the results of our GALFIT modeling (\S\ref{sec:galfit}), the AGN sample remains significantly offset from the \citet{Kormendy2013} relation (see the Appendix). We find that the median horizontal bulge offset when using GALFIT is -1.40 $\pm$ 0.70 when including a central point source and -1.22 $\pm$ 0.75 dex when not including a central point source in the models. The median horizontal offset does not change substantially between GALFIT and GIM2D, or between including/not including a central point source.

For comparison, we also find the median offsets between the groups of additional objects described in \S\ref{sec:Additional Objects} and the \citet{Kormendy2013} BH-bulge mass relation. Figures \ref{fig:KH_shifts} shows the results for both total stellar mass and bulge mass; extracting the bulge component always brings the galaxies closer to the canonical relation.  As expected, the elliptical galaxies along with classical bulges in spiral galaxies originally used to derive the canonical \citet{Kormendy2013} relation have a median horizontal and vertical offset of approximately zero. However, even with the bulge extracted, all the other types of galaxies tend to fall below the relation to varying degrees. Reverberation-mapped AGNs and dwarf AGNs/composites fall $\sim$0.5 dex  below the relation on average. Megamaser galaxies and pseudobulges fall even further below the relation, at $\sim$0.7 dex. Our sample of low-luminosity, broad-line AGNs falls the furthest (more than 1 dex) below the canonical relation, with zero offset inconsistent within {$2\sigma$} from the median. We note that some of these objects have small sample sizes in Figure \ref{fig:KH_shifts}.

\subsection{Linear Fit}
\label{sec:fit}

{We fit a linear relation between log BH mass and log stellar bulge mass, as well as log total stellar mass for our sample. For consistency, we follow the procedure outlined by \citet{Reines2015} using the Bayesian linear regression method of \citet{Kelly2007} which takes into account errors in both the BH and bulge or total stellar mass. We parameterize the relation with the following:}

\begin{equation}
    \mathrm{log(M_{BH}/M_\odot) = \alpha + \beta \: log(M_{\star}/10^{11}M_\odot})
\end{equation}

\noindent
and find,

\begin{equation}
\begin{gathered}
     \mathrm{\alpha_{Bulge} = 7.18 \pm 0.09, \: \beta_{Bulge} = 0.60 \pm 0.11} \\
     \mathrm{\alpha_{Total} = 7.03 \pm 0.07, \: \beta_{Total} = 0.72 \pm 0.13},
\end{gathered}
\end{equation}

\noindent    
where the slope and intercept are the median of 10,000 draws from the posterior probability distribution of the parameters and the reported errors are the 1$\sigma$ scatter. 

{We compare the BH-bulge mass relation best-fit parameters that we find for our low-luminosity, broad-line AGNs to values in the literature derived from various samples of galaxies (shown in Figure \ref{fig:linregress}). We include the \citet{Kormendy2013} relation as well as the  \citet{Haring2004} and \citet{McConnell2013} relations, all of which are based on samples of early-type galaxies. We also include the relation derived by \citet{Schutte2019} based on a variety of galaxy types and included here as a comparison sample of additional objects (see \S\ref{sec:Additional Objects}). \citet{Schutte2019} include 12 dwarf galaxies \citep{Reines2013}, 37 reverberation-mapped AGN from \citet{Bentz2018}, nine megamaser galaxies from \citet{Lasker2016} and the \citet{Kormendy2013} sample.}

We also show four relations in the literature between BH mass and total stellar mass in Figure \ref{fig:linregress}. We first include the AGN relation from \citet{Reines2015}. Although our sample is a subset of their sample, we find a different BH mass to total stellar mass relation with a shallower slope. There are likely two reasons for this. First, we retain less than half of their original sample after cross-matching with \citet{Simard2011} and removing galaxies with high AGN-to-bulge luminosity ratios that could lead to unreliable bulge-disk decompositions (see \S\ref{sec:simard_justification}). Second, our total stellar masses are calculated somewhat differently as described in \S\ref{sec:mass_calculations}. {\bf For completeness, we also include in Figure \ref{fig:linregress} the BH mass to total stellar mass relation in \citet{Reines2015} for early-type galaxies with quiescent BHs.}

We also include the two relations found by \citet{Greene2016}. They combine a sample of seven megamaser galaxies with the \citet{saglia2016} sample of 97 elliptical, classical bulge and pseudobulge galaxies and then split this sample into early-type and late-type galaxies to calculate a regression line relating BH-to-total stellar mass for each separately. Our slope again is shallower and our intercept is lower than both of these comparison relations. 

The variety of scaling relations derived here and from the literature shown in Figure \ref{fig:linregress} demonstrates the criticality of using the most relevant relation for a given galaxy sample. The various relations apply to different properties of galaxies (i.e., $M_{\rm BH}$ vs.\ $M_{\rm \star,Bulge}$ \: $\neq$ \: $M_{\rm BH}$ vs.\ $M_{\rm \star,Total}$) and/or are derived from samples biased towards certain types of galaxies. As a result, there are significantly different slopes and intercepts among the various relations. As with previous works, we emphasize that all galaxies do not follow a single all-encompassing scaling relation between BH mass and either bulge or total stellar mass. Moreover, it is important to continue to investigate and derive scaling relations for diverse types of galaxies in order to estimate BH masses for similar samples.

\subsection{No Strong Correlations with Galaxy Properties}
\label{sec:galaxy_properties}

To explore possible causes for deviations from the canonical BH-bulge relation for our sample, we examine correlations between offset in BH mass from the \citet{Kormendy2013} relation (i.e., how far off the observed BH mass is from the value predicted by the canonical relation at a given bulge mass) as a function of various potentially relevant galaxy properties. Figure \ref{fig:galaxy_properties} shows the offset in BH mass versus bulge $g-r$ color, \dn\ strength\footnote{The \dn\ parameter \citep{Balogh1999} describes the average flux ratio of the continuum on the red versus blue side of the 4000\textup{~\AA} break. A higher value indicates a relatively weak UV/blue component in the spectrum due to a lack of young hot stars and increased absorption in the atmospheres of older, more metal rich stars. A \dn\ value of 1.5 corresponds to an average stellar population of $\sim$ 1 Gyr and therefore is generally used as a criteria for separating star-forming and quiescent galaxies \citep{Kauffmann2003a,Kauffmann2003b,Hathi2009,Haines2017,Kim2018}. We use the \dn\ strength as reported in the NSA, which is based on spectroscopy obtained within a 3\arcsec\ fiber.}, AGN $r$-band luminosity, AGN-to-bulge $r$-band luminosity ratio, bulge and galaxy Sérsic indices, redshift, and bulge-to-total $r$-band luminosity. We do not find a strong correlation between the offset from the \citet{Kormendy2013} relation with any of these properties (see Figure \ref{fig:galaxy_properties}).

We calculate the Pearson and Spearman rank correlation coefficients ($r$) to quantitatively assess the strength of the correlation between each of these galaxy properties with the offset from the \citet{Kormendy2013} relation. The Spearman rank correlation coefficient describes how well the relationship can be described by a monotonic function and the Pearson correlation describes how well it can be described specifically by a linear function. We find moderate correlations ($0.4 \gtrsim r \gtrsim 0.6$) for the offset from the \citet{Kormendy2013} relation versus $g-r$ color and AGN fraction but the rest of the galaxy properties show weak correlations ($r \lesssim 0.4$). The coefficients are shown in Figure \ref{fig:galaxy_properties}. All of these {values} are statistically significant based on their p-values calculated using a t-test. 

The moderate negative correlation between offset in BH mass with bulge color indicates that bluer bulges tend to lie closer to the \citet{Kormendy2013} relation while the reddest bulges fall the farthest below the relation. This result could be interpreted to support a connection between BH growth and star formation, either via concurrent fueling and/or AGN feedback. However, we caution that this result could also be an artifact from the color dependency of the mass-to-light ratios. That is, for a given luminosity, bulges that are redder in color receive a higher stellar mass estimate.

\section{Conclusion and Discussion}\label{sec:conclusion}

In this paper we test whether or not the canonical BH-bulge mass relation holds for a sample of 117 nearby ($z \le 0.055$), low-luminosity, broad-line AGNs identified by \citet{Reines2015} with bulge-disk decompositions from \citet{Simard2011}. Investigating active galaxies probes a subset of galaxies that is distinct from samples of early-type galaxies with dynamically detected BHs that have been used to derive canonical BH mass - bulge mass relations \citep[e.g.,][]{Kormendy2013}. 

Our work was primarily motivated by the findings of \citet{Reines2015} regarding BH mass - total stellar mass relations in the local Universe. In particular, they found that their sample of broad-line AGNs fell on average more than an order of magnitude below early-type galaxies with quiescent BHs. Our main goal here was to determine whether or not this same sample of broad-line AGNs follows the canonical BH-to-bulge mass relation. 

We find that our sample of active galaxies as a whole tends to fall below canonical scaling relations, hosting central BHs that are smaller than predicted by the BH mass $-$ bulge mass relation by a factor of $\sim$10. We do not find a strong systematic trend between any galaxy property with offset from the the canonical relation.

Whether or not dwarf galaxies follow the canonical scaling relations may indicate signatures of BH seeding mechanisms, however, observations in this mass regime are scarce. While \citet{Schutte2019} and \citet{Baldassare2020} find that their samples of dwarf galaxies align with the canonical $M_{\rm BH}-M_{\rm bulge}$ and $M_{\rm BH}-\sigma$ relations, respectively, both of these works note the selection bias for finding brighter BHs, which are typically of higher mass, meaning that they could simply be detecting the upper envelope of BH masses in dwarf galaxies and potentially the rest of the population lies below the canonical relations.

Our work adds to the growing body of evidence that the canonical BH-to-bulge scaling relations derived from highly biased samples of early-type galaxies with dynamical BH mass measurements should be used with caution as they are not applicable to all types of galaxies \citep{Wandel1999, Greene2008, Gadotti2009,Greene2010, Kormendy2011}. In their work, \citet{Kormendy2013} note that this relation only holds for classical bulges and ellipticals and should not be applied to objects such as pseudobulges, which tend to fall below the relation. Through this work, we emphasize the importance of using the most appropriate scaling relation according to the properties of a given sample, since we find that our sample of local, low-luminosity, broad-line AGNs are significantly offset from the canonical BH-bulge mass scaling relation defined by early-type galaxies. Consequently, this work has important implications for predicting black hole merger rates for the {\it Laser Interferometer Space Antenna} and Pulsar Timing Arrays that will detect the gravitational wave emission from these events, as well as cosmological simulations that are tied to the local BH-to-bulge mass relations. 
 
\vspace{1cm}
We thank the referee for their time and providing a helpful report that improved that overall quality of this manuscript.
AER acknowledges support provided by NASA through EPSCoR grant number 80NSSC20M0231 and the NSF through CAREER award 2235277.

\appendix

\begin{figure*}
    \begin{center}
    \includegraphics[width=5.5in]{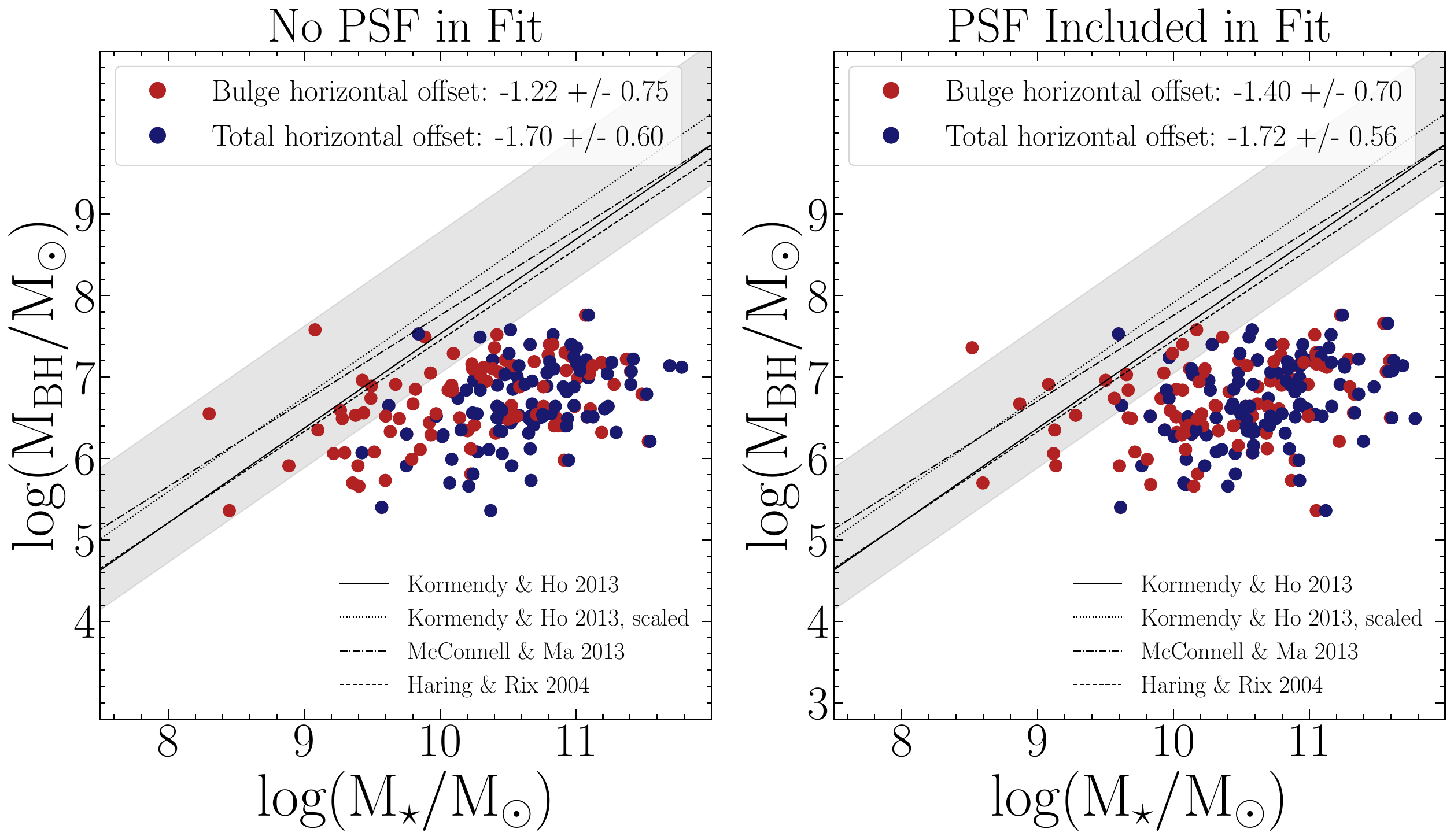}
    \caption{BH mass versus total stellar mass (blue) and stellar bulge mass (red) using decompositions from GALFIT (see \S\ref{sec:galfit}). The left panel shows results from decompositions done using a two component bulge$+$disk model, while the right panel shows results from decompositions done using a three component bulge$+$disk$+$central point source (PSF) model. Both using a two and three component model results in our sample being offset from the canonical relations. The \citet{Kormendy2013}, \cite{Haring2004}, and \citet{McConnell2013} BH-bulge mass relations are shown, with the shaded region indicating the 3$\sigma$ intrinsic scatter of the \citet{Kormendy2013} relation as reported in their paper.}
    \label{fig:galfit_mbh_mgal}
        \end{center}
\end{figure*}

We show the BH-bulge mass relation calculated using our GALFIT decompositions in Figure \ref{fig:galfit_mbh_mgal}. Our sample of low-mass AGNs remains offset from canonical relations even if we use the GALFIT models where we perform three component decompositions using a bulge$+$disk$+$central point source, and in fact, the offset is greater than when using the two component decompositions done by \citet{Simard2011}.

\begin{figure*}
\centering
    \includegraphics[width=5.5in]{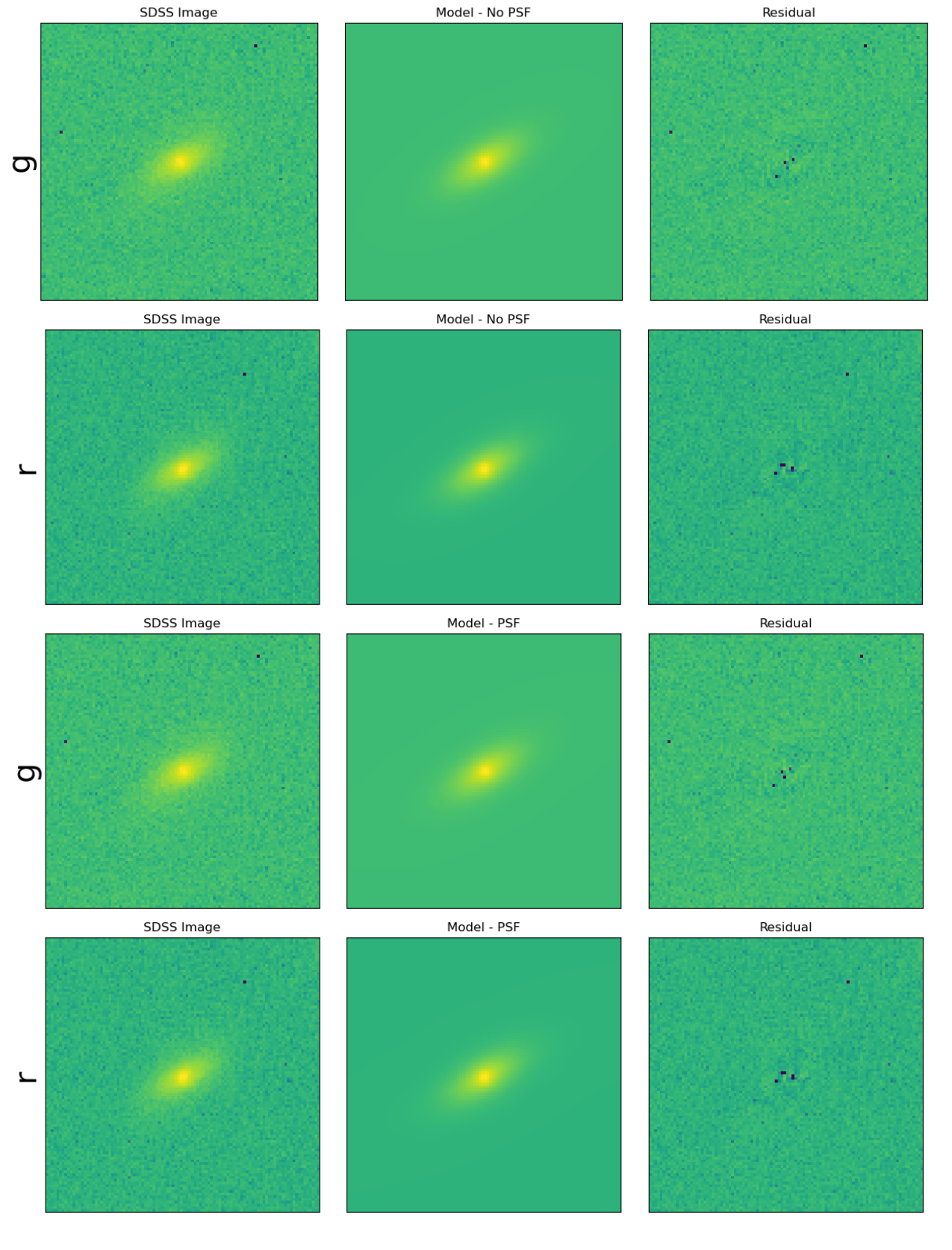}
    \caption{An example of the GALFIT models made for one galaxy (ID 587725469598417019 from \citet{Simard2011}). The columns show the SDSS images (left), the GALFIT models (middle) and the resulting residuals (right). The top two rows show fits using a two component (bulge$+$disk) model in both g- and r-bands and the bottom two rows show fits using a three component (bulge$+$disk$+$PSF) model in both g- and r-bands. }
    \label{fig:galfit_model_example}
\end{figure*}

We show an example of the GALFIT decompositions that we performed for one galaxy in Figure \ref{fig:galfit_model_example}. We perform these fits using both two and three component models in each band. The resulting galaxy component masses that we find are very similar to those calculated by \citet{Simard2011} using two component decompositions.

\newpage
\bibliography{mybib}

\begin{thebibliography}{}
\expandafter\ifx\csname natexlab\endcsname\relax\def\natexlab#1{#1}\fi
\providecommand{\url}[1]{\href{#1}{#1}}
\providecommand{\dodoi}[1]{doi:~\href{http://doi.org/#1}{\nolinkurl{#1}}}
\providecommand{\doeprint}[1]{\href{http://ascl.net/#1}{\nolinkurl{http://ascl.net/#1}}}
\providecommand{\doarXiv}[1]{\href{https://arxiv.org/abs/#1}{\nolinkurl{https://arxiv.org/abs/#1}}}

\bibitem[{Aihara {et~al.}(2011)Aihara, Prieto, An, Anderson, Aubourg, Balbinot, Beers, Berlind, Bickerton, Bizyaev, Blanton, Bochanski, Bolton, Bovy, Brandt, Brinkmann, Brown, Brownstein, Busca, Campbell, Carr, Chen, Chiappini, Comparat, Connolly, Cortes, Croft, Cuesta, da~Costa, Davenport, Dawson, Dhital, Ealet, Ebelke, Edmondson, Eisenstein, Escoffier, Esposito, Evans, Fan, Castell{\'{a}}, Font-Ribera, Frinchaboy, Ge, Gillespie, Gilmore, Hern{\'{a}}ndez, Gott, Gould, Grebel, Gunn, Hamilton, Harding, Harris, Hawley, Hearty, Ho, Hogg, Holtzman, Honscheid, Inada, Ivans, Jiang, Johnson, Jordan, Jordan, Kazin, Kirkby, Klaene, Knapp, Kneib, Kochanek, Koesterke, Kollmeier, Kron, Lampeitl, Lang, Goff, Lee, Lin, Long, Loomis, Lucatello, Lundgren, Lupton, Ma, MacDonald, Mahadevan, Maia, Makler, Malanushenko, Malanushenko, Mandelbaum, Maraston, Margala, Masters, McBride, McGehee, McGreer, M{\'{e}}nard, Miralda-Escud{\'{e}}, Morrison, Mullally, Muna, Munn, Murayama, Myers, Naugle, Neto, Nguyen, Nichol,
  O{\textquotesingle}Connell, Ogando, Olmstead, Oravetz, Padmanabhan, Palanque-Delabrouille, Pan, Pandey, P{\^{a}}ris, Percival, Petitjean, Pfaffenberger, Pforr, Phleps, Pichon, Pieri, Prada, Price-Whelan, Raddick, Ramos, Reyl{\'{e}}, Rich, Richards, Rix, Robin, Rocha-Pinto, Rockosi, Roe, Rollinde, Ross, Ross, Rossetto, S{\'{a}}nchez, Sayres, Schlegel, Schlesinger, Schmidt, Schneider, Sheldon, Shu, Simmerer, Simmons, Sivarani, Snedden, Sobeck, Steinmetz, Strauss, Szalay, Tanaka, Thakar, Thomas, Tinker, Tofflemire, Tojeiro, Tremonti, Vandenberg, Maga{\~{n}}a, Verde, Vogt, Wake, Wang, Weaver, Weinberg, White, White, Yanny, Yasuda, Yeche, \& Zehavi}]{Aihara2011}
Aihara, H., Prieto, C.~A., An, D., {et~al.} 2011, The Astrophysical Journal Supplement Series, 193, 29, \dodoi{10.1088/0067-0049/193/2/29}

\bibitem[{{Baldassare} {et~al.}(2020){Baldassare}, {Dickey}, {Geha}, \& {Reines}}]{Baldassare2020}
{Baldassare}, V.~F., {Dickey}, C., {Geha}, M., \& {Reines}, A.~E. 2020, \apjl, 898, L3, \dodoi{10.3847/2041-8213/aba0c1}

\bibitem[{{Baldassare} {et~al.}(2016){Baldassare}, {Reines}, {Gallo}, {Greene}, {Graur}, {Geha}, {Hainline}, {Carroll}, \& {Hickox}}]{Baldassare2016}
{Baldassare}, V.~F., {Reines}, A.~E., {Gallo}, E., {et~al.} 2016, \apj, 829, 57, \dodoi{10.3847/0004-637X/829/1/57}

\bibitem[{{Balogh} {et~al.}(1999){Balogh}, {Morris}, {Yee}, {Carlberg}, \& {Ellingson}}]{Balogh1999}
{Balogh}, M.~L., {Morris}, S.~L., {Yee}, H.~K.~C., {Carlberg}, R.~G., \& {Ellingson}, E. 1999, \apj, 527, 54, \dodoi{10.1086/308056}

\bibitem[{{Begelman} {et~al.}(2006){Begelman}, {Volonteri}, \& {Rees}}]{Begelman2006}
{Begelman}, M.~C., {Volonteri}, M., \& {Rees}, M.~J. 2006, \mnras, 370, 289, \dodoi{10.1111/j.1365-2966.2006.10467.x}

\bibitem[{Bell {et~al.}(2003)Bell, McIntosh, Katz, \& Weinberg}]{Bell2003}
Bell, E.~F., McIntosh, D.~H., Katz, N., \& Weinberg, M.~D. 2003, The Astrophysical Journal Supplement Series, 149, 289–312, \dodoi{10.1086/378847}

\bibitem[{{Bellovary} {et~al.}(2019){Bellovary}, {Cleary}, {Munshi}, {Tremmel}, {Christensen}, {Brooks}, \& {Quinn}}]{Bellovary2019}
{Bellovary}, J.~M., {Cleary}, C.~E., {Munshi}, F., {et~al.} 2019, \mnras, 482, 2913, \dodoi{10.1093/mnras/sty2842}

\bibitem[{{Bentz} \& {Manne-Nicholas}(2018)}]{Bentz2018}
{Bentz}, M.~C., \& {Manne-Nicholas}, E. 2018, \apj, 864, 146, \dodoi{10.3847/1538-4357/aad808}

\bibitem[{Blanton {et~al.}(2011)Blanton, Kazin, Muna, Weaver, \& Price-Whelan}]{Blanton2011}
Blanton, M.~R., Kazin, E., Muna, D., Weaver, B.~A., \& Price-Whelan, A. 2011, The Astronomical Journal, 142, 31, \dodoi{10.1088/0004-6256/142/1/31}

\bibitem[{{Blanton} {et~al.}(2005){Blanton}, {Schlegel}, {Strauss}, {Brinkmann}, {Finkbeiner}, {Fukugita}, {Gunn}, {Hogg}, {Ivezi{\'c}}, {Knapp}, {Lupton}, {Munn}, {Schneider}, {Tegmark}, \& {Zehavi}}]{Blanton2005}
{Blanton}, M.~R., {Schlegel}, D.~J., {Strauss}, M.~A., {et~al.} 2005, \aj, 129, 2562, \dodoi{10.1086/429803}

\bibitem[{{Bond} {et~al.}(1984){Bond}, {Arnett}, \& {Carr}}]{Bond1984}
{Bond}, J.~R., {Arnett}, W.~D., \& {Carr}, B.~J. 1984, \apj, 280, 825, \dodoi{10.1086/162057}

\bibitem[{{Conroy} {et~al.}(2009){Conroy}, {Gunn}, \& {White}}]{Conroy2009}
{Conroy}, C., {Gunn}, J.~E., \& {White}, M. 2009, \apj, 699, 486, \dodoi{10.1088/0004-637X/699/1/486}

\bibitem[{Gadotti \& Kauffmann(2009)}]{Gadotti2009}
Gadotti, D.~A., \& Kauffmann, G. 2009, Monthly Notices of the Royal Astronomical Society, 399, 621, \dodoi{10.1111/j.1365-2966.2009.15328.x}

\bibitem[{Gebhardt {et~al.}(2000)Gebhardt, Bender, Bower, Dressler, Faber, Filippenko, Green, Grillmair, Ho, Kormendy, Lauer, Magorrian, Pinkney, Richstone, \& Tremaine}]{Gebhardt2000}
Gebhardt, K., Bender, R., Bower, G., {et~al.} 2000, The Astrophysical Journal, 539, L13, \dodoi{10.1086/312840}

\bibitem[{Getachew-Woreta {et~al.}(2022)Getachew-Woreta, Povi{\'{c} }, Masegosa, Perea, Beyoro-Amado, \& M{\'{a}}rquez}]{Getachew_Woreta2022}
Getachew-Woreta, T., Povi{\'{c} }, M., Masegosa, J., {et~al.} 2022, Monthly Notices of the Royal Astronomical Society, 514, 607, \dodoi{10.1093/mnras/stac851}

\bibitem[{{Ghosh} {et~al.}(2023){Ghosh}, {Urry}, {Mishra}, {Perreault-Levasseur}, {Natarajan}, {Sanders}, {Nagai}, {Tian}, {Cappelluti}, {Kartaltepe}, {Powell}, {Rau}, \& {Treister}}]{Ghosh2023}
{Ghosh}, A., {Urry}, C.~M., {Mishra}, A., {et~al.} 2023, \apj, 953, 134, \dodoi{10.3847/1538-4357/acd546}

\bibitem[{Graham \& Scott(2014)}]{Graham2014}
Graham, A.~W., \& Scott, N. 2014, The Astrophysical Journal, 798, 54, \dodoi{10.1088/0004-637x/798/1/54}

\bibitem[{Greene {et~al.}(2008)Greene, Ho, \& Barth}]{Greene2008}
Greene, J.~E., Ho, L.~C., \& Barth, A.~J. 2008, The Astrophysical Journal, 688, 159, \dodoi{10.1086/592078}

\bibitem[{Greene {et~al.}(2010)Greene, Peng, Kim, Kuo, Braatz, Impellizzeri, Condon, Lo, Henkel, \& Reid}]{Greene2010}
Greene, J.~E., Peng, C.~Y., Kim, M., {et~al.} 2010, The Astrophysical Journal, 721, 26, \dodoi{10.1088/0004-637x/721/1/26}

\bibitem[{Greene {et~al.}(2016)Greene, Seth, Kim, Läsker, Goulding, Gao, Braatz, Henkel, Condon, Lo, \& Zhao}]{Greene2016}
Greene, J.~E., Seth, A., Kim, M., {et~al.} 2016, The Astrophysical Journal, 826, L32, \dodoi{10.3847/2041-8205/826/2/l32}

\bibitem[{{Haines} {et~al.}(2017){Haines}, {Iovino}, {Krywult}, {Guzzo}, {Davidzon}, {Bolzonella}, {Garilli}, {Scodeggio}, {Granett}, {de la Torre}, {De Lucia}, {Abbas}, {Adami}, {Arnouts}, {Bottini}, {Cappi}, {Cucciati}, {Franzetti}, {Fritz}, {Gargiulo}, {Le Brun}, {Le F{\`e}vre}, {Maccagni}, {Ma{\l}ek}, {Marulli}, {Moutard}, {Polletta}, {Pollo}, {Tasca}, {Tojeiro}, {Vergani}, {Zanichelli}, {Zamorani}, {Bel}, {Branchini}, {Coupon}, {Ilbert}, {Moscardini}, {Peacock}, \& {Siudek}}]{Haines2017}
{Haines}, C.~P., {Iovino}, A., {Krywult}, J., {et~al.} 2017, \aap, 605, A4, \dodoi{10.1051/0004-6361/201630118}

\bibitem[{{Hathi} {et~al.}(2009){Hathi}, {Ferreras}, {Pasquali}, {Malhotra}, {Rhoads}, {Pirzkal}, {Windhorst}, \& {Xu}}]{Hathi2009}
{Hathi}, N.~P., {Ferreras}, I., {Pasquali}, A., {et~al.} 2009, \apj, 690, 1866, \dodoi{10.1088/0004-637X/690/2/1866}

\bibitem[{Hu(2008)}]{Hu2008}
Hu, J. 2008, Monthly Notices of the Royal Astronomical Society, 386, 2242, \dodoi{10.1111/j.1365-2966.2008.13195.x}

\bibitem[{{Huertas-Company} {et~al.}(2011){Huertas-Company}, {Aguerri}, {Bernardi}, {Mei}, \& {S{\'a}nchez Almeida}}]{Huertas2011}
{Huertas-Company}, M., {Aguerri}, J.~A.~L., {Bernardi}, M., {Mei}, S., \& {S{\'a}nchez Almeida}, J. 2011, \aap, 525, A157, \dodoi{10.1051/0004-6361/201015735}

\bibitem[{Häring \& Rix(2004)}]{Haring2004}
Häring, N., \& Rix, H.-W. 2004, The Astrophysical Journal, 604, L89, \dodoi{10.1086/383567}

\bibitem[{{Kauffmann} {et~al.}(2003{\natexlab{a}}){Kauffmann}, {Heckman}, {White}, {Charlot}, {Tremonti}, {Brinchmann}, {Bruzual}, {Peng}, {Seibert}, {Bernardi}, {Blanton}, {Brinkmann}, {Castander}, {Cs{\'a}bai}, {Fukugita}, {Ivezic}, {Munn}, {Nichol}, {Padmanabhan}, {Thakar}, {Weinberg}, \& {York}}]{Kauffmann2003a}
{Kauffmann}, G., {Heckman}, T.~M., {White}, S. D.~M., {et~al.} 2003{\natexlab{a}}, \mnras, 341, 33, \dodoi{10.1046/j.1365-8711.2003.06291.x}

\bibitem[{{Kauffmann} {et~al.}(2003{\natexlab{b}}){Kauffmann}, {Heckman}, {White}, {Charlot}, {Tremonti}, {Peng}, {Seibert}, {Brinkmann}, {Nichol}, {SubbaRao}, \& {York}}]{Kauffmann2003b}
---. 2003{\natexlab{b}}, \mnras, 341, 54, \dodoi{10.1046/j.1365-8711.2003.06292.x}

\bibitem[{Kelly(2007)}]{Kelly2007}
Kelly, B.~C. 2007, The Astrophysical Journal, 665, 1489, \dodoi{10.1086/519947}

\bibitem[{{Kim} {et~al.}(2018){Kim}, {Malhotra}, {Rhoads}, {Joshi}, {Fererras}, \& {Pasquali}}]{Kim2018}
{Kim}, K., {Malhotra}, S., {Rhoads}, J.~E., {et~al.} 2018, arXiv e-prints, arXiv:1810.01498, \dodoi{10.48550/arXiv.1810.01498}

\bibitem[{Kormendy {et~al.}(2011)Kormendy, Bender, \& Cornell}]{Kormendy2011}
Kormendy, J., Bender, R., \& Cornell, M.~E. 2011, Nature, 469, 374, \dodoi{10.1038/nature09694}

\bibitem[{{Kormendy} \& {Ho}(2013)}]{Kormendy2013}
{Kormendy}, J., \& {Ho}, L.~C. 2013, \araa, 51, 511, \dodoi{10.1146/annurev-astro-082708-101811}

\bibitem[{{Loeb} \& {Rasio}(1994)}]{Loeb1994}
{Loeb}, A., \& {Rasio}, F.~A. 1994, \apj, 432, 52, \dodoi{10.1086/174548}

\bibitem[{Läsker {et~al.}(2016)Läsker, Greene, Seth, van~de Ven, Braatz, Henkel, \& Lo}]{Lasker2016}
Läsker, R., Greene, J.~E., Seth, A., {et~al.} 2016, The Astrophysical Journal, 825, 3, \dodoi{10.3847/0004-637x/825/1/3}

\bibitem[{Magorrian {et~al.}(1998)Magorrian, Tremaine, Richstone, Bender, Bower, Dressler, Faber, Gebhardt, Green, Grillmair, Kormendy, \& Lauer}]{Magorrian1998}
Magorrian, J., Tremaine, S., Richstone, D., {et~al.} 1998, The Astronomical Journal, 115, 2285, \dodoi{10.1086/300353}

\bibitem[{Marconi \& Hunt(2003)}]{Marconi2003}
Marconi, A., \& Hunt, L.~K. 2003, The Astrophysical Journal, 589, L21, \dodoi{10.1086/375804}

\bibitem[{McConnell \& Ma(2013)}]{McConnell2013}
McConnell, N.~J., \& Ma, C.-P. 2013, The Astrophysical Journal, 764, 184, \dodoi{10.1088/0004-637x/764/2/184}

\bibitem[{{Mendel} {et~al.}(2014){Mendel}, {Simard}, {Palmer}, {Ellison}, \& {Patton}}]{Mendel2014}
{Mendel}, J.~T., {Simard}, L., {Palmer}, M., {Ellison}, S.~L., \& {Patton}, D.~R. 2014, \apjs, 210, 3, \dodoi{10.1088/0067-0049/210/1/3}

\bibitem[{Merritt \& Ferrarese(2001)}]{Merritt2001}
Merritt, D., \& Ferrarese, L. 2001, The Astrophysical Journal, 547, 140, \dodoi{10.1086/318372}

\bibitem[{{Peng} {et~al.}(2010){Peng}, {Ho}, {Impey}, \& {Rix}}]{Peng2010}
{Peng}, C.~Y., {Ho}, L.~C., {Impey}, C.~D., \& {Rix}, H.-W. 2010, \aj, 139, 2097, \dodoi{10.1088/0004-6256/139/6/2097}

\bibitem[{Reines {et~al.}(2013)Reines, Greene, \& Geha}]{Reines2013}
Reines, A.~E., Greene, J.~E., \& Geha, M. 2013, The Astrophysical Journal, 775, 116, \dodoi{10.1088/0004-637x/775/2/116}

\bibitem[{Reines \& Volonteri(2015)}]{Reines2015}
Reines, A.~E., \& Volonteri, M. 2015, The Astrophysical Journal, 813, 82, \dodoi{10.1088/0004-637x/813/2/82}

\bibitem[{{Ricarte} \& {Natarajan}(2018)}]{Ricarte2018}
{Ricarte}, A., \& {Natarajan}, P. 2018, \mnras, 481, 3278, \dodoi{10.1093/mnras/sty2448}

\bibitem[{{Saglia} {et~al.}(2016){Saglia}, {Opitsch}, {Erwin}, {Thomas}, {Beifiori}, {Fabricius}, {Mazzalay}, {Nowak}, {Rusli}, \& {Bender}}]{saglia2016}
{Saglia}, R.~P., {Opitsch}, M., {Erwin}, P., {et~al.} 2016, \apj, 818, 47, \dodoi{10.3847/0004-637X/818/1/47}

\bibitem[{Sani {et~al.}(2011)Sani, Marconi, Hunt, \& Risaliti}]{Sani2011}
Sani, E., Marconi, A., Hunt, L.~K., \& Risaliti, G. 2011, Monthly Notices of the Royal Astronomical Society, 413, 1479, \dodoi{10.1111/j.1365-2966.2011.18229.x}

\bibitem[{{Schutte} {et~al.}(2019){Schutte}, {Reines}, \& {Greene}}]{Schutte2019}
{Schutte}, Z., {Reines}, A.~E., \& {Greene}, J.~E. 2019, \apj, 887, 245, \dodoi{10.3847/1538-4357/ab35dd}

\bibitem[{Simard {et~al.}(2011)Simard, Trevor~Mendel, Patton, Ellison, \& McConnachie}]{Simard2011}
Simard, L., Trevor~Mendel, J., Patton, D.~R., Ellison, S.~L., \& McConnachie, A.~W. 2011, The Astrophysical Journal Supplement Series, 196, 11, \dodoi{10.1088/0067-0049/196/1/11}

\bibitem[{Simard {et~al.}(2002)Simard, Willmer, Vogt, Sarajedini, Phillips, Weiner, Koo, Im, Illingworth, \& Faber}]{Simard2002}
Simard, L., Willmer, C. N.~A., Vogt, N.~P., {et~al.} 2002, The Astrophysical Journal Supplement Series, 142, 1–33, \dodoi{10.1086/341399}

\bibitem[{{Volonteri}(2010)}]{Volonteri2010}
{Volonteri}, M. 2010, \aapr, 18, 279, \dodoi{10.1007/s00159-010-0029-x}

\bibitem[{{Volonteri} {et~al.}(2008){Volonteri}, {Lodato}, \& {Natarajan}}]{Volonteri2008}
{Volonteri}, M., {Lodato}, G., \& {Natarajan}, P. 2008, \mnras, 383, 1079, \dodoi{10.1111/j.1365-2966.2007.12589.x}

\bibitem[{{Volonteri} \& {Natarajan}(2009)}]{Volonteri2009}
{Volonteri}, M., \& {Natarajan}, P. 2009, \mnras, 400, 1911, \dodoi{10.1111/j.1365-2966.2009.15577.x}

\bibitem[{{Wandel}(1999)}]{Wandel1999}
{Wandel}, A. 1999, \apjl, 519, L39, \dodoi{10.1086/312106}

\bibitem[{{Wang} {et~al.}(2021){Wang}, {Yang}, {Fan}, {Hennawi}, {Barth}, {Banados}, {Bian}, {Boutsia}, {Connor}, {Davies}, {Decarli}, {Eilers}, {Farina}, {Green}, {Jiang}, {Li}, {Mazzucchelli}, {Nanni}, {Schindler}, {Venemans}, {Walter}, {Wu}, \& {Yue}}]{Feige2021}
{Wang}, F., {Yang}, J., {Fan}, X., {et~al.} 2021, \apjl, 907, L1, \dodoi{10.3847/2041-8213/abd8c6}

\bibitem[{{Willick} {et~al.}(1997){Willick}, {Courteau}, {Faber}, {Burstein}, {Dekel}, \& {Strauss}}]{Willick1997}
{Willick}, J.~A., {Courteau}, S., {Faber}, S.~M., {et~al.} 1997, \apjs, 109, 333, \dodoi{10.1086/312983}

\bibitem[{Yan(2011)}]{Yan2011}
Yan, R. 2011, The Astronomical Journal, 142, 153, \dodoi{10.1088/0004-6256/142/5/153}

\bibitem[{Yan \& Blanton(2012)}]{Yan2012}
Yan, R., \& Blanton, M.~R. 2012, The Astrophysical Journal, 747, 61, \dodoi{10.1088/0004-637x/747/1/61}

\bibitem[{Zibetti {et~al.}(2009)Zibetti, Charlot, \& Rix}]{Zibetti2009}
Zibetti, S., Charlot, S., \& Rix, H.-W. 2009, Monthly Notices of the Royal Astronomical Society, 400, 1181–1198, \dodoi{10.1111/j.1365-2966.2009.15528.x}

\end{thebibliography}

\end{document}